\documentclass[11pt]{article}

\usepackage{newtxtext,newtxmath}

\usepackage{graphicx}

\usepackage[letterpaper,margin=1in]{geometry}

\renewenvironment{abstract}
	{\quotation}
	{\endquotation}

\date{}

\makeatletter
\renewcommand{\fnum@figure}{\textbf{Figure \thefigure}}
\renewcommand{\fnum@table}{\textbf{Table \thetable}}
\makeatother

\usepackage{scicite}

\usepackage{url}

\usepackage{soul}
\usepackage{xcolor}
\sethlcolor{yellow}
\newif\ifshowchanges
\showchangestrue   
\newcommand{\hlchange}[1]{#1}

\usepackage{caption}

\DeclareCaptionFont{ninept}{\fontsize{9}{11}\selectfont}
\def\scititle{
	Predicting upcoming visual features during eye movements yields scene representations aligned with human visual cortex
}
\title{\bfseries \boldmath \scititle}

\author{
	Sushrut Thorat$^{1,4\ast}$,
	Adrien Doerig$^{1,2,3}$,
	Alexander Kroner$^{1}$,\and
    Carmen Amme$^{1}$,
    Tim C. Kietzmann$^{1\ast}$\and
	\small$^{1}$Institute of Cognitive Science, Osnabrück University, Osnabrück 49069, Germany.\and
	\small$^{2}$Department of Education and Psychology, Freie Universität, Berlin 14195, Germany.\and
    \small$^{3}$Bernstein Center for Computational Neuroscience, Berlin 10115, Germany.\and
	\small$^{4}$Lossfunk, Bengaluru 560038, India.\,\,\,\,\,\,\,\,\,\,\,\,\,\,\,\,\,\,\,\,\,\, \and
	\small$^\ast$Corresponding authors. Emails: sthorat@uos.de, tkietzma@uos.de\and
}

\begin{document} 

\maketitle

\noindent\textbf{Short title:} Predicting glimpses yields brain-like scene codes

\noindent\textbf{Teaser:} Predicting upcoming visual features helps neural networks learn brain-like scene representations.

\begin{abstract} \bfseries \boldmath
Natural scenes are complex arrangements of objects, surfaces, and backgrounds. For the brain’s visual system to effectively operate, it needs to extract not only what objects are present, but also their spatial and semantic relations. We hypothesize that such structures may be learned, in a self-supervised fashion, by exploiting temporal regularities of natural active vision: each fixation reveals a glimpse that is related to the previous one via co-occurrence and saccade-conditioned spatial regularities. We instantiate this idea with Glimpse Prediction Networks (GPNs), recurrent models trained to predict the embedding of the next glimpse along human-like scanpaths. GPNs are shown to successfully extract complex scene information, including object co-occurrences and spatial object arrangements, and integrate information across glimpses. Importantly, GPN representations align strongly with human fMRI responses in mid and higher-level visual cortex and match, often outperform, alternative state-of-the-art ANN models, establishing next-glimpse-prediction as a biologically plausible route towards brain-aligned scene representations.
\end{abstract}


\noindent
The advent of neuroconnectionism in the past decade delivered image-computable models of visual-cortex representations of natural object images~\cite{doerig_neuroconnectionist_2023,richards_deep_2019}. While most prior success was reported for representations of single (natural) objects~\cite{yamins_using_2016}, natural scenes pose a tougher challenge. Scenes contain multiple objects, occlusions, surfaces, and backgrounds, all of which are in spatial and semantic relations with each other~\cite{peelen_category_2017,epstein_scene_2019}. What approach could yield neural networks that are able to unify these scene parts and their relations into brain-aligned scene representations? 

\hlchange{As standard object-classification objectives do not explicitly supervise such relations, in search for an alternative,} recent studies turned to large-language model (LLM) embeddings of scene captions. These embeddings contain information of namable scene parts and their relations, which were used as a supervised learning target. Networks trained with this objective acquired representations that aligned better with scene representations across visual cortex than object-classification networks~\cite{wang_better_2023,doerig_high-level_2025}. Yet, two issues with this approach remain. First, the LLM embeddings are themselves based on large amounts of training data, raising questions about their biological plausibility~\cite{frank_bridging_2023}. Second, LLM embeddings were also shown to align with scene representations of macaques, who do not have access to complex language~\cite{conwell_monkey_2025}. These observations raise the question of how such rich, brain-aligned scene representations could be learned without language.

We propose that an answer lies in the information available to the brain during active vision. Humans explore the world through sequences of fixations, with each fixation delivering a high-resolution glimpse of a part of the visual scene in front of us~\cite{yarbus_eye_1967}. Sequences of glimpses are rich sources of information. For example they encode co-occurrence statistics of scene parts (e.g., tiles → sink is likely; tiles → whale is not)~\cite{kaiser_object_2019} as well as information about their spatial arrangement, conveyed by saccadic efference copies (e.g., from a car, an upward eye movement tends to reveal sky, not road)~\cite{summerfield_structure_2020}. To induce these statistics into neural networks, we take inspiration from computational work on structure discovery in sequences~\cite{elman_finding_1990}: we propose that predicting the visual features of the upcoming glimpse could serve as a meaningful objective, as it requires knowing what scene parts could occur next given the saccade. Additionally, recurrence could help build a unified scene representation: executing sequential predictions may benefit from the ability to integrate information over glimpses to improve future predictions. Following this logic, we hypothesized that predicting the next glimpse features with a recurrent architecture may constitute an objective that would lead to rich, unified, scene representations.

\begin{figure}[!t] 
	\centering
	\includegraphics[width=0.8\textwidth]{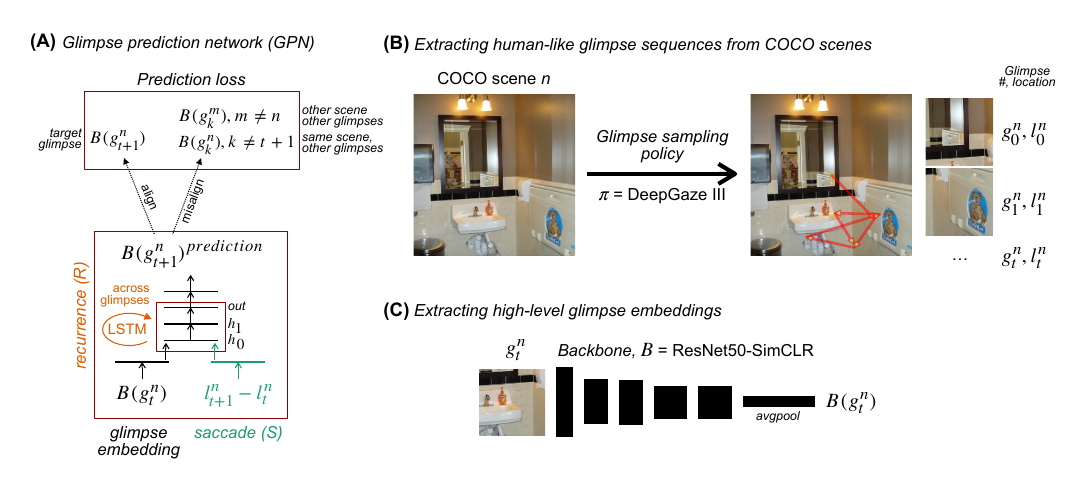} 

	\caption{\textbf{Glimpse Prediction Networks (GPNs) for scene understanding.} (\textbf{A}) GPNs took glimpse embeddings, and predicted the next glimpse embeddings in a sequence, with optional access to saccades (S) and recurrence (R). They were trained with a contrastive prediction loss. We trained the four GPN variants — GPN-B (no saccades, no recurrence), GPN-R, GPN-S, and GPN-RS (saccades and recurrence) — and assessed their behavior and the alignment of their internal representations with scene representations in the human visual cortex.
	(\textbf{B}) The glimpse centers were sampled from human-like eye-movement sequences over 256px COCO scenes, modeled with DeepGazeIII. 91px glimpses were cropped around these centers.
	(\textbf{C}) The glimpse embeddings were extracted with pretrained backbones to encode high-level visual features. For example, the primary results were obtained using the ResNet50-SimCLR backbone, which is a ResNet50 trained with the SimCLR self-supervised learning objective on the ILSVRC dataset. The \emph{avgpool} representation of the glimpse was used as its embedding. These backbones were frozen during GPN training.}
	\label{fig:img-net-depiction} 
\end{figure}

To test this hypothesis, we designed Glimpse Prediction Networks (GPNs): recurrent deep neural networks trained to predict the post-saccadic features—operationalized as high-level embeddings of glimpse pixels from pretrained networks—in human-like glimpse sequences over natural-scene images (Fig.~\ref{fig:img-net-depiction}A). Model variants include GPNs with/without recurrence and with/without saccadic information. We report that GPNs are able to acquire knowledge of feature co-occurrence and, when saccade information is made available, spatial arrangement of natural scene parts. With recurrence, they can infer and integrate such relationships across parts in novel scenes. Importantly, the GPN parameters were trained entirely through self-supervised next-glimpse prediction, without scene labels, language supervision, or neural data; the visual backbones and DeepGazeIII sampling model were pretrained and frozen. The resulting internal representations were found to align well with scene representations in human visual cortex, better than data/parameter-matched networks supervised to predict caption embeddings, and state-of-the-art vision and vision-language neural networks. These findings highlight a biologically meaningful path towards self-supervised models of human scene representation.

\subsection*{Results}

Glimpse Prediction Networks (GPNs) are trained with a contrastive objective~\cite{oord_representation_2019,bakhtiari_functional_2021} to predict the (high-level) embedding of the next saccade target (glimpse), given a sequence of prior fixations and saccadic efferent copies (inter-glimpse displacements). That is, the generated prediction should be closer (in cosine similarity terms) to the embedding of the next glimpse and farther from the embeddings of other glimpses from the same scene or other scenes. To better understand the impact of the various ingredients, GPN variants were created using a $2\times 2$ design with the factors recurrence (presence denoted as R, implemented as an LSTM or its non-recurrent equivalent), and saccade target location (presence denoted with S), implemented as a $2$D displacement vector from the center of the given glimpse to the center of the next glimpse (Fig.~\ref{fig:img-net-depiction}A). The model variant with neither recurrence nor saccadic efferent copies was denoted B (baseline). Human-like input glimpse sequences were modeled using DeepGazeIII~\cite{kummerer_deepgaze_2022}, over Common Objects in Context (COCO) scene images~\cite{lin_microsoft_2014} (Fig.~\ref{fig:img-net-depiction}B). \hlchange{We used DeepGazeIII because it predicts scanpaths, not just static saliency: given an image and fixation history, it predicts a distribution over the next fixation location, from which human-like glimpse sequences can be sampled. It also reported state-of-the-art performance on free-viewing scanpath prediction at the time. Reliance of the results reported below (specifically, the neural-alignment results) on the components of this glimpse-sampling policy is depicted in Fig.}~\ref{fig:saccadic-policy-results}. To obtain a high-level embedding of the glimpses, we used a self-supervised, SimCLR-trained ResNet50 (Fig.~\ref{fig:img-net-depiction}C), but note that similar results can be obtained with other supervised and self-supervised backbones (RN50-ILSVRC and DINOv2-B; see Figs.~\ref{fig:backbone-2-examples} \& \ref{fig:backbone-2-behavior}).

\subsubsection*{GPNs learn the spatial co-occurrence of scene parts and integrate across them}

We first asked whether next-glimpse prediction produces three key properties of scene representations: sensitivity to which scene parts co-occur, sensitivity to their spatial arrangement, and integration of information across successive glimpses.

\begin{figure}[!t]
	\centering
	\includegraphics[width=0.8\textwidth]{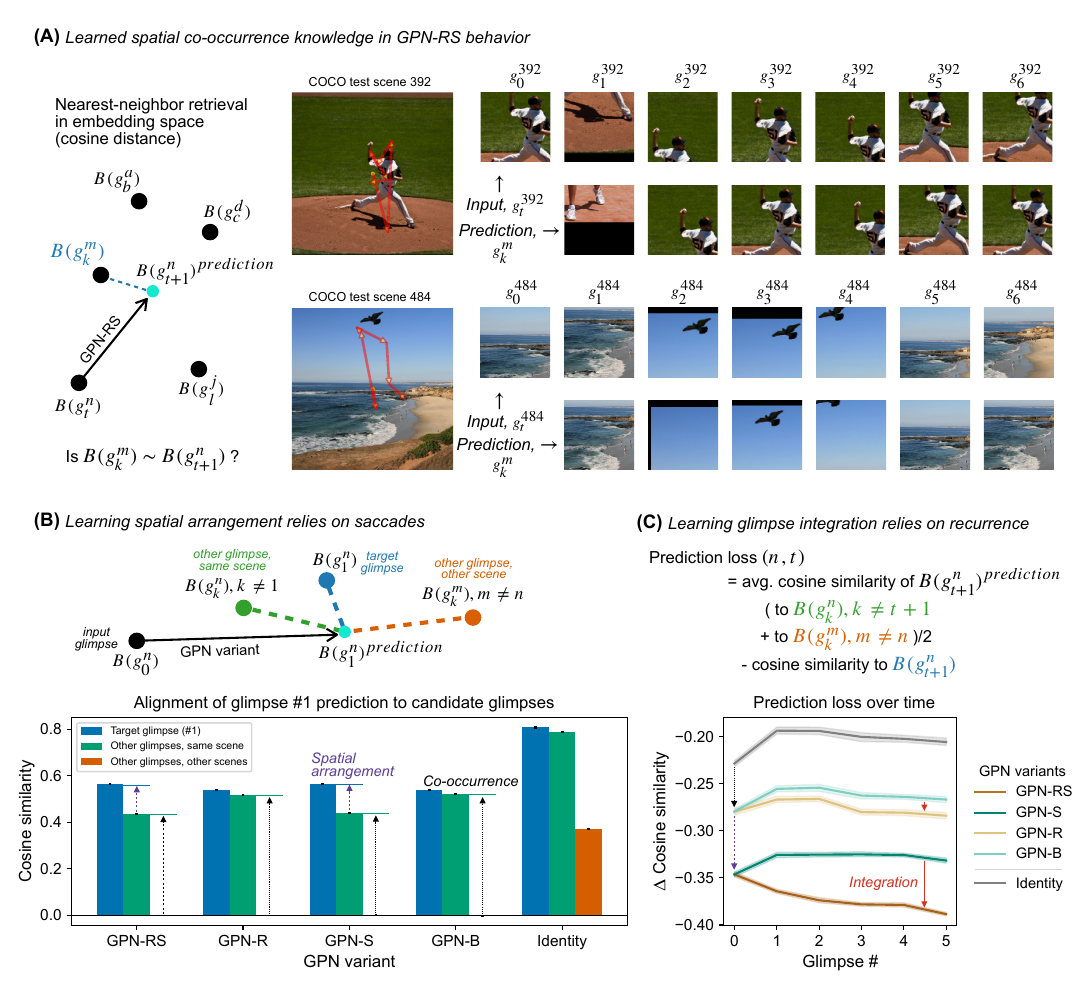}

	\caption{\textbf{GPNs learn co-occurrence, spatial arrangement, and integration of scene parts.}
(\textbf{A}) Qualitative examples of GPN-RS predictions on held-out COCO scenes. Given an input glimpse embedding and a saccade, GPN-RS predicted an embedding whose nearest neighbor often corresponded to plausible upcoming scene content. For example, a downward saccade from a baseball-player glimpse led to predictions aligned with a lower-body and ground glimpse; in another scene, upward and downward saccades led to predictions aligned with sky/bird and beach glimpses, respectively. Nearest-neighbor retrievals were computed in backbone embedding space using cosine distance.
(\textbf{B}) GPN predictions captured co-occurrence and spatial-arrangement statistics. Predictions from the first, central, glimpse ($g_0$) were compared with the target next glimpse, other glimpses from the same scene, and glimpses from other scenes. All GPNs' predictions aligned more strongly with same-scene glimpses than with other-scene glimpses, indicating learned co-occurrence structure. GPN-S and GPN-RS, which had access to saccades, had predictions aligned more strongly with the target next glimpse than with other same-scene glimpses, indicating sensitivity to spatial arrangement.
(\textbf{C}) Recurrence supported integration across glimpses. Prediction loss was computed as the average cosine similarity to non-target glimpses minus the cosine similarity to the target next glimpse, such that lower values indicate better prediction. Recurrent GPNs showed increasingly lower prediction loss over the glimpse sequence than their non-recurrent counterparts, consistent with memory-based integration across glimpses. Error bars indicate 95\% CIs.}
	\label{fig:beh-analysis} 
\end{figure}

We qualitatively assessed the behavior of GPN-RS, the variant with the lowest prediction loss, on held-out COCO scenes (`test-515'). As seen in Fig.~\ref{fig:beh-analysis}A (examples across variants in Fig.~\ref{fig:backbone-2-examples}), \hlchange{an input embedding of a glimpse containing a baseball player's upper body, accompanied by a saccade towards the bottom, led the GPN to predict an embedding that aligned best with a glimpse containing the feet of a sportsperson and the ground (across all glimpses from all held-out scenes). In another scene, a saccade towards the top from a glimpse (embedding) depicting a beach led the GPN to predict an embedding that aligned best with a glimpse depicting the sky. After receiving the subsequent glimpse input (embedding), depicting a bird in the sky, its subsequent embedding predictions corrected, aligning best with glimpses containing a bird. Additionally, a saccade towards the bottom led to an embedding prediction corresponding to the beach glimpse seen before.} These examples suggest GPN-RS successfully learned about the co-occurrence and spatial arrangement of glimpse features in natural scenes.

To quantify this learning of glimpse co-occurrence, due to training for glimpse predictions, we compared the alignment of the GPN predictions, given the first glimpse ($g_0$), with other glimpses from the same scene and with glimpses from the other scenes. Indeed, paired tests revealed a higher alignment for other glimpses from the same scene than glimpses from other scenes for all GPN variants (mean difference of cosines, $\bar{\Delta} > 0.41$, Wilcoxon signed-rank tests: $W = 0$, FDR-adjusted $p = 5.9\times 10^{-8}$; $N = 25$ batches of $206$ glimpse sequences each; Fig.~\ref{fig:beh-analysis}B). Additionally, the cosine similarity of the GPNs’ prediction to the embedding of the input glimpse was much lower than $1$ ($1-\mu > 0.34$, Wilcoxon signed-rank tests: $W = 0$, FDR-adjusted $p = 5.9\times 10^{-8}$), showing that the GPNs were not trivially implementing an Identity mapping. Finally, the prediction’s similarity to embeddings of glimpses from other scenes was close to zero ($\mu < 0.002$ for the GPNs, which was expected to be $\mu = 0.37$ \hlchange{assuming the ‘Identity’ mapping—constituting a baseline condition revealing effects expected given the cosine similarities amongst the input glimpse embeddings, by setting the GPN output = glimpse embedding input}). These results suggest GPNs learned a transformation that effectively orthogonalized glimpses from different scenes while maintaining similarity between glimpses within scenes, a sign that they accounted for the co-occurrence of glimpses.

To assess the dependence of learning of glimpse spatial arrangements on saccadic information, we compared the alignment of the GPN predictions, given the first glimpse ($g_0$), with the next, target, glimpse and with other glimpses from the same scene. \hlchange{If the cosine similarity of the GPN prediction with the target glimpse embedding is higher than its similarity with other glimpse embeddings from the same scene, then spatial arrangement is taken into account by the GPN.} Checking across GPN variants, a $2\times 2$ (recurrence $\times$ saccade presence) within-subject nonparametric contrast analysis revealed a main effect of saccades on the aforementioned difference between cosine similarities ($W = 0$, Holm-adjusted $p = 1.7\times 10^{-7}$; no interaction with recurrence, $W = 152$, Holm-adjusted $p = 0.79$; and no main effect of recurrence, $W = 97$, Holm-adjusted $p = 0.16$; Fig.~\ref{fig:beh-analysis}B). After accounting for (i.e., subtracting) the cosine similarities expected from the Identity baseline, we found better alignment with the next glimpse than other glimpses from the same scene for GPN-RS and GPN-S (averaged differences across variants; mean difference of cosines, $\bar{\Delta} = 0.11$, $W = 0$, FDR-adjusted $p = 1.2\times 10^{-7}$) but not for GPN-R and GPN-B ($\bar{\Delta} = 4.6\times 10^{-4}$, $W = 161$, FDR-adjusted $p = 0.98$). These results suggest that, with access to saccades, GPNs successfully learned a transformation over glimpse embeddings, accounting for the spatial arrangement of glimpses in scenes.

An important design aspect towards learning unified scene representations was the inclusion of recurrence, which was expected to enable GPNs to integrate information across glimpses in service of improving future glimpse predictions. Starting with the GPNs without recurrence, GPN-B \& GPN-S, we observed that the prediction loss varied with glimpse number (Friedman tests, $\chi^2_{24} \geq 59$, FDR-adjusted $p = 1.3\times 10^{-11}$) and was lowest for the first, central, glimpse (reflecting glimpse embedding statistics, captured by the Identity condition). The addition of recurrence resulted in a prediction loss that was lower compared to the corresponding non-recurrent GPNs, an effect that became increasingly evident with increasing glimpse number (Wilcoxon signed-rank comparisons over loss difference correlations with glimpse number; GPN-RS vs GPN-S: mean $\bar{\Delta} = 0.79$, $W = 0$, FDR-adjusted $p = 1.1\times 10^{-7}$; GPN-R vs GPN-B: $\bar{\Delta} = 0.39$, $W = 37$, FDR-adjusted $p = 3.3\times 10^{-4}$). These results suggest that, with access to recurrence, GPNs learned to use their memory states to drive information integration across glimpses in scenes, in the service of improving future glimpse embedding predictions.


To further characterize how GPN-RS integrates information across glimpses, we tested whether it learned a situated, content-location memory rather than relying only on natural spatial co-occurrence statistics. First, we trained cross-validated linear decoders to predict each glimpse's absolute x,y location from model representations across layers and glimpse numbers (Fig.~\ref{fig:gpn-rs-binding}A). In GPN-RS, location was decoded near ceiling from the first LSTM layer and remained available as glimpses accumulated, whereas this pattern was absent before training and in models lacking either recurrence or saccade inputs. This suggests that GPN-RS learned to path integrate the saccades and maintain a location-indexed recurrent state. We next tested whether this state could bind arbitrary content to locations by assigning glimpse embeddings from different scenes to fixation positions, thereby breaking natural co-occurrence and spatial-arrangement statistics, and simulating return saccades to previously visited locations (Fig.~\ref{fig:gpn-rs-binding}B). GPN-RS retrieved the embedding previously encountered at the target location even for these artificial sequences, including when the embedding dimensions were independently shuffled. Thus, GPN-RS learned to construct a situated representation of what appeared where within the current sequence, rather than merely exploiting learned scene statistics. Because these arbitrary content-location bindings were formed online without weight updates, this ability can be viewed as a specific form of in-context learning—namely, in-context content-location binding~\cite{brown_language_2020}.

Taken together, our analyses suggest that GPNs learned transformations that contextualized the glimpse embeddings by including: (i) across scenes, information about co-occurrence, and spatial arrangement (given saccades), of scene parts, and (ii) integrated information across glimpses seen in the current scene (given recurrence). Based on these observations, we consider that the sequences of glimpse embeddings were transformed into more holistic scene representations. 

\subsubsection*{GPN internal representations align with scene representations in visual cortex}

Having established that GPNs encode co-occurrence, spatial arrangement, and information accumulated across glimpses, we next asked whether these learned representations explain cortical scene representations better than the input glimpse features, matched-objective control networks, and strong alternative models.

\begin{figure}[!t]
	\centering
	\includegraphics[width=0.8\textwidth]{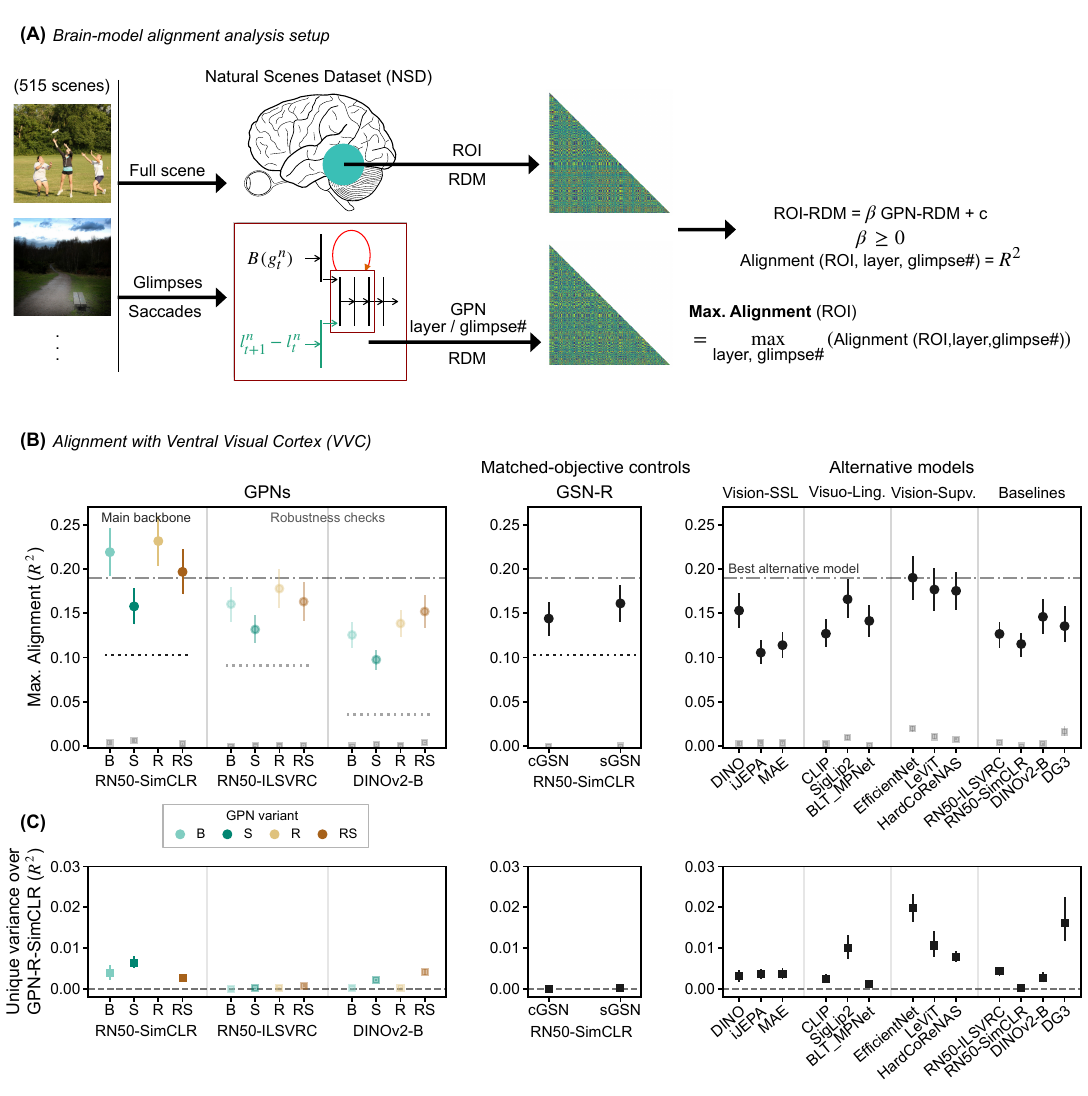}

	\caption{\textbf{GPNs exhibit improved representational alignment with the scene representations in ventral visual cortex (VVC), surpassing state-of-the-art models.}
	(\textbf{A}) Representational dissimilarity matrices (RDMs) were computed from scene representations in a given ROI and the considered network. Alignment was quantified as the variance of ROI RDM explained by a network RDM via non-negative linear regression. As the networks have multiple layers (and glimpse numbers in the case of GPNs) we report the maximum alignment across those layer/glimpse\# comparisons per network. 
	(\textbf{B}) Maximum alignment with the Ventral Visual Cortex (VVC), of the considered networks, is shown.
	(\textbf{left}) All GPNs, across variants and backbones, had higher alignment than their input glimpse embeddings (dotted black lines). Amongst the GPNs, GPN-R with the RN50-SimCLR backbone had the highest VVC-alignment. Overall, recurrence helped alignment. Saccades hurt alignment over GPN-B. However, over GPN-R, they had mixed influence (as further described in Fig.~\ref{fig:gpn-backbones-sweep}). 
	(\textbf{middle}) GPN-R alignment was higher than Glimpse Stitching Networks (GSNs) trained with the same architecture but different objectives: to predict the classes of objects in the scenes (cGSN) or semantic embeddings of scene captions (sGSN).
	(\textbf{right}) A large number of state-of-the-art models had lower VVC-alignment than the RN50-SimCLR-based GPN-B and GPN-R (highest alignment across those models, for EfficientNet-B3, is indicated by the black line).
	(\textbf{C}) Moreover, other models explained little to no unique variance in the VVC RDMs over that explained by the RN50-SimCLR-based GPN-R (also indicated by square markers in (B)). Error bars indicate 95\% CIs.}
	\label{fig:main-neural-results} 
\end{figure}

Scene representations were obtained from the Natural Scenes Dataset (NSD)~\cite{allen_massive_2022}. We considered the special-515 COCO scenes that all participants observed. Importantly, GPNs were not trained on these scenes. To assess the alignment of the internal representations of the GPNs with representations in the visual cortex, we conducted representational similarity analyses (RSA)~\cite{kriegeskorte_representational_2008}. For each scene, we used one sampled glimpse sequence. \hlchange{Across scenes, for a given glimpse number ($t$) in the sequence and a layer of a given GPN, including the input glimpse embedding as the first representational stage, we computed a representational dissimilarity matrix (RDM) across the 515 scenes and compared it with an RDM of a given region-of-interest (ROI) in the brain.} Thus, each layer-by-glimpse-number pair yielded one GPN RDM for comparison with the corresponding ROI RDM. Alignment was assessed via non-negative linear regression, as shown in Fig.~\ref{fig:main-neural-results}A. \hlchange{For each GPN--ROI comparison, we summarized alignment as the maximum variance explained across layers and glimpse numbers. The same procedure was used for comparison networks, except that these networks had no glimpse-number dimension. The selected layers and glimpse numbers for the maximal alignment scores are reported in Tables}~\ref{tab:model-family-overview}--\ref{tab:vision-language-vvc}. We primarily focus on the alignment of GPN representations to the representations observed in the high-level ventral stream ROI, which subsumes high-level object and scene areas, such as the PPA~\cite{epstein_cortical_1998} (`ventral' ROI from the NSD streams definition; hereafter termed ventral visual cortex, VVC).


We started by characterizing the influence of the various GPN components (saccades, recurrence, and backbones) on VVC-alignment. For this, we computed the \hlchange{\textit{maximum}} alignment score for each GPN variant and three backbones (RN50-SimCLR, RN50-ILSVRC, and DINOv2-B) across all glimpse numbers in the sequences, and layers (Fig.~\ref{fig:main-neural-results}A; detailed results in Tables~\ref{tab:model-family-overview}--\ref{tab:vision-language-vvc}). The degree of VVC-alignment depended on an interaction of the inclusion of recurrence and saccades, and the backbone ($2\times 2\times 3$ repeated-measures analysis; $3$-way interaction: $\chi^2_{2} = 14.3$, $p = 8\times 10^{-4}$; Fig.~\ref{fig:main-neural-results}B, upper-left panel).
\hlchange{The addition of recurrence to GPN-B and GPN-S increased their VVC-alignment} (main effect; mean difference, $\bar{\Delta} = 0.028$, $W = 0$, Holm-adjusted $p = 0.02$). 
Interestingly, the addition of saccades to GPN-B variants reduced their VVC-alignment (averaged across backbones; $\bar{\Delta} = -0.039$, $W = 0$, Holm-adjusted $p = 0.0078$). However, the addition of saccades to GPN-R did not reliably increase or decrease the VVC-alignment (decrease for RN50-SimCLR/ILSVRC: $\bar{\Delta} < -0.0147$, $W = 0$, FDR-adjusted $p = 0.0078$; increase for DINOv2-B: $\bar{\Delta} = 0.013$, $W = 0$, FDR-adjusted $p = 0.0078$). We note, for the best performing backbone, RN50-SimCLR, this negative influence of saccades was found throughout the brain, as indicated by a searchlight analysis that revealed \hlchange{GPN-R to be in better alignment than GPN-RS} \hlchange{across the whole brain} (Fig.~\ref{fig:searchlight-analysis}). \hlchange{A sweep across other well-known ResNet, Convolutional and Transformer backbones supported these observations: recurrence helped VVC-alignment, but saccades hurt VVC-alignment over GPN-B, and the relationship between the addition of saccades to GPN-R and VVC-alignment was not systematic} (see Fig.~\ref{fig:gpn-backbones-sweep}).

Importantly, for each GPN variant and backbone, the best alignment score was higher than the maximum alignment of their input glimpse embeddings (denoted by black dotted lines in Fig.~\ref{fig:main-neural-results}B, upper-left panel; across all GPNs, $\bar{\Delta} > 0.04$, $W = 0$, FDR-adjusted $p = 0.0078$). These results suggest that the GPNs, taking the co-occurrence of scene parts into account and integrating information across them, did transform glimpse embeddings into scene-like representations aligned with representations of full natural scenes in the ventral visual cortex.

Next, to assess if the glimpse prediction objective—as opposed to the dataset (\hlchange{NSD images were also taken from COCO, which could lead to better alignment than the non-COCO trained backbone embeddings}) and architecture (additional layers and recurrence, which could have increased expressivity)—was essential for achieving the increased degree of representational alignment with VVC, \hlchange{we considered alternative objectives}, keeping the architecture and dataset matched to the most VVC-aligned, GPN-R with the RN50-SimCLR backbone. \hlchange{We term the resulting networks Glimpse Stitching Networks (GSNs), which incorporate two previously-studied objectives}~\cite{doerig_high-level_2025}. Classification-GSN (cGSN) was trained to categorize the objects in the scene, and Semantic-GSN (sGSN) was trained to predict the semantic embeddings (MPNet-based) of the corresponding scene captions. As seen in Fig.~\ref{fig:main-neural-results}B (middle panel), both the GSN-R variants exhibited lower VVC-alignment than GPN-R (cGSN: mean difference, $\bar{\Delta} = 0.088$, $W = 0$, FDR-adjusted $p = 0.0078$; sGSN: $\bar{\Delta} = 0.075$, $W = 0$, FDR-adjusted $p = 0.0078$). This result highlights the utility of the glimpse prediction objective, rather than the dataset or architecture changes, in acquiring scene representations aligned with the ventral visual cortex.

To assess how the VVC-alignment of the GPNs compared to related and state-of-the-art approaches, we benchmarked VVC-alignment across a large number of alternative models. First, we considered self-supervised (SSL) vision network families that were trained with objectives that fostered learning part-whole relationships by aligning representations across image parts (DINO, iJEPA, MAE)~\cite{caron_emerging_2021,assran_self-supervised_2023,he_masked_2021}. Second, we considered visuo-linguistic network families that were trained to align visual representations of scenes with representations of their corresponding captions (CLIP, SigLiP2, BLT-MPNet)~\cite{doerig_high-level_2025,radford_learning_2021,tschannen_siglip_2025}. Third, we considered network families with state-of-the-art alignment with occipitotemporal cortex~\cite{conwell_large-scale_2024} (EfficientNet, LeViT, HardCoReNAS)~\cite{tan_efficientnet_2020,graham_levit_2021,nayman_hardcore-nas_2021}. Fourth, we considered the three backbone variants used for GPN (RN50-ILSVRC, RN50-SimCLR, and DINOv2-B) and DeepGazeIII (the generation of human-like fixations could involve human-like scene representation), as strong baselines. For each family of networks, we report the best observed alignment score across all networks and their corresponding layers in those families (Fig.~\ref{fig:main-neural-results}B, upper-right panel). To account for the possibility that a central glimpse (approximating foveal vision, equivalent to glimpse $0$ for the GPNs) might provide better alignment than the full scene (the typical input for these networks), given that the participants fixated centrally in the NSD experiment, we furthermore report the maximum alignment across the full scene vs central glimpse comparisons for all alternative models.

\hlchange{GPN-R with the RN50-SimCLR backbone} achieved the highest VVC alignment amongst all tested networks (Fig.~\ref{fig:main-neural-results}B, upper panels). Amongst all alternative models tested, EfficientNet (B3 variant, trained for image classification on the ILSVRC dataset), presented with the central glimpse, was best aligned with VVC scene representations (alignment results for all the networks tested can be seen summarized in Table~\ref{tab:model-family-overview} and detailed in Tables~\ref{tab:gpn-gsn-vvc}--\ref{tab:vision-language-vvc}). Amongst the self-supervised networks, a DINO-trained network (DINOv2-ViT-L14 variant) was the best-aligned. Yet, \hlchange{GPN-B and GPN-R with the RN50-SimCLR backbone} had significantly higher alignment than both of these highest-performing alternative models (mean differences across comparisons of GPN variants and backbones with the two best networks, $\bar{\Delta} > 0.028$, $W = 36$, FDR-adjusted $p = 0.016$). In sum, the scene representations emerging due to the glimpse prediction objective, given the RN50-SimCLR backbone, surpassed the state-of-the-art models of scene representations in the ventral visual cortex. 

Although GPN-R, with the RN50-SimCLR backbone, explained the most variance in VVC amongst all the networks tested, it could have been the case that it explained different variance to the alternative models, as the objectives, architectures, and training datasets vary substantially. To assess this possibility, we conducted variance-partitioning analyses, to assess if the alternative models explain any unique VVC-variance beyond that explained by GPN-R. As seen in Fig.~\ref{fig:main-neural-results}C, the highest unique variance explained by any other network (over GPN-R), was observed for EfficientNet ($\mu = 0.019$, $W = 0$, $p = 0.007$). Yet, GPN-R accounted for $90\%$ of variance explained by EfficientNet. The unique variance explained by the other networks was much lower, and we did not find evidence for the dataset/parameter-matched cGSN and sGSN explaining any unique variance ($\mu < 4\times 10^{-4}$, $W = 0$, FDR-adjusted $p = 0.33$). In sum, GPN-R, with the RN50-SimCLR backbone, largely subsumed the variance explained in VVC by alternative, large-scale networks, which had access to the full scene images (during training and test) and in some cases access to rich semantic information through scene captions.

To test whether the high GPN alignment is specific to VVC, we extended our analysis to the other NSD \textit{streams} ROIs: early, midventral, midlateral, lateral, midparietal, and parietal. First, for GPN-R, with the RN50-SimCLR backbone, the increase in alignment over input glimpse embeddings held across these ROIs ($\bar{\Delta} > 0.012$, $W = 0$, FDR-adjusted $p = 0.0078$; Fig.~\ref{fig:comparison-allROIs}). Second, the GPN-R showed higher alignment than cGSN and sGSN across these ROIs (vs. max GSN alignments, $\bar{\Delta} > 0.0058$, $W < 2$, FDR-adjusted $p < 0.015$). Third, relative to the best-aligned alternative model for each ROI, the GPN-R was better-aligned with the midparietal and parietal ROIs ($\bar{\Delta} > 0.01$, $W < 3$, FDR-adjusted $p < 0.05$), similarly-aligned with midventral, midlateral, and lateral ROIs ($\bar{\Delta} > -0.0075$, $W > 3$, FDR-adjusted $p > 0.082$), and worse-aligned only with the early ROI ($\bar{\Delta} = -0.06$, $W = 0$, FDR-adjusted $p = 0.023$)—understandably, as the alternative models could better capture low-level features, whereas GPNs begin with high-level glimpse embeddings. Finally, in midparietal and parietal ROIs, where GPN-R surpassed alternatives, it explained $\geq88\%$ of the total variance explained by any other model. These results indicate that the glimpse-prediction objective yields scene representations aligning with large swaths of mid/high-level visual cortex, outperforming parameter/dataset-matched controls and matching or exceeding strong large-scale baselines, subsuming most of the variance they explain.

\begin{figure}[!t]
	\centering
	\includegraphics[width=0.6\textwidth]{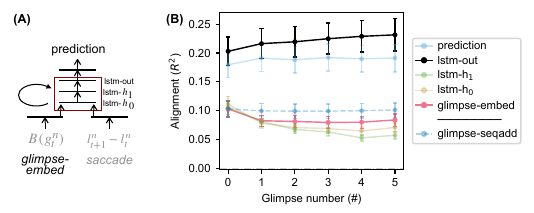}

	\caption{\textbf{GPN-R transforms glimpse embeddings into a VVC-aligned recurrent state.}
(\textbf{A}) Layer-wise analysis of the RN50-SimCLR-based GPN-R. We computed representational alignment with ventral visual cortex (VVC) for the input glimpse embedding, recurrent hidden states, LSTM output, prediction layer, and a sequential averaging baseline. Alignment was quantified as the variance of the VVC RDM explained by each model RDM via non-negative linear regression.
(\textbf{B}) The LSTM output showed the strongest VVC-alignment. Input glimpse embeddings explained a portion of VVC representational structure, but the LSTM output substantially increased alignment from the first glimpse onward and improved further across the glimpse sequence. The \emph{glimpse-seqadd} baseline was computed by sequentially averaging glimpse-embedding representations across the sequence, testing whether simple accumulation of viewed glimpse features was sufficient. LSTM-output alignment exceeded this baseline, indicating that recurrence did more than average the input glimpse embeddings. The prediction layer showed lower alignment than the LSTM output, suggesting that the most cortex-aligned representation emerges in the recurrent state optimized to support next-glimpse prediction. Error bars indicate 95\% CIs.}
	\label{fig:gpn-r-simclr-internals} 
\end{figure}

Having observed that \hlchange{GPN-R, with the RN50-SimCLR backbone}, trained in a self-supervised fashion, achieved state-of-the-art alignment with ventral stream scene representation in NSD, we asked which layer, and hence computation, drives this effect. As seen in Fig.~\ref{fig:gpn-r-simclr-internals}, the RDMs of the input glimpse embeddings (glimpse-embed) accounted for $\sim 10\%$ of the variance of VVC RDMs. For the first (central) glimpse, the LSTM output (lstm-out) nearly doubled this alignment (mean increase $\Delta = 0.099$, $W = 0$, $N = 8$, $p = 0.0078$). With additional glimpses, due to recurrence, the alignment of lstm-out increased further (mean Spearman correlation between the alignment gain and glimpse number, $\rho = 0.97$, $W = 0$, $p = 0.0078$). Notably, this increase in alignment was higher than expected by sequential averaging of glimpse embedding representations (glimpse-seqadd; reminiscent of building sentence embeddings from word embeddings; comparison of the maximum alignment of `lstm-out' vs `glimpse-seqadd': $\bar{\Delta} = 0.127$, $W = 0$, $p = 0.0078$). Critically, the best alignment was found at the LSTM output layer, which outperformed both the input and the prediction layers (e.g., at glimpse $\#5$, the prediction layer alignment was lower than lstm-out; $\Delta = 0.04$, $W = 0$, $p = 0.0078$). Intermediate layers did not show comparable gains. We observed such maximum VVC-alignment in the LSTM output layer across all the GPN variants and backbones (Fig.~\ref{fig:gpn-internals-all}; also for the midparietal/parietal ROIs, see Fig.~\ref{fig:comparison-midparietal-parietal}). In sum, GPNs transform input glimpse embeddings into a mid/high-level visual-cortex-aligned internal state, optimized for predicting the next-glimpse embedding. 

\subsection*{Discussion}

By taking stock of information available to the primate visual system, here we presented a self-supervised learning objective that predicts a high-level embedding of the next-glimpse in a sequence of human-like eye-movements. This approach, termed Glimpse Prediction Networks (GPNs), resulted in networks that successfully internalize the co-occurrence of scene parts and their spatial arrangement, and that integrate information across glimpses into a rich scene representation. Remarkably, these emergent representations outperformed their input-glimpse features in their alignment with scene representations across visual cortex, as well as Glimpse Stitching Networks trained to predict scene objects or caption embeddings. In mid/high-level visual cortex, these emergent representations equated or outperformed previously reported state-of-the-art ventral stream models, including large-scale foundation models.

The most intriguing finding of the current study was that the variance in ventral visual cortex (VVC; as well as midparietal/parietal) scene representations explained by diverse models—supervised, self-supervised, vision-only, and visuo-linguistic—was largely subsumed by that explained by the best-aligned GPN (GPN-R, with a ResNet50-SimCLR backbone). This suggests that the shared variance chiefly hinges on what this GPN-R captured: co-occurrence of scene parts and an integration across them. This convergence suggests a common structure for cortical scene representation alignment across current models~\cite{huh_position_2024}. Notably, the results suggest that previously reported LLM–brain alignment~\cite{wang_better_2023,doerig_high-level_2025} might be mediated primarily by co-occurrence structure — the same signals that underpin distributional semantics in language models that emerge through a similar objective: next-token prediction~\cite{harris_distributional_1954,mikolov_efficient_2013,brown_language_2020}.

An unexpected result was that adding the saccade signal often reduced GPN alignment with NSD scene representations. This was clearest in the non-recurrent models, where GPN-S aligned worse than GPN-B across nearly all backbones, but it also occurred for some recurrent models. This is surprising because the same saccade information improved next-glimpse prediction, suggesting that better task performance does not necessarily imply better cortical alignment. Ventral stream responses are sensitive to object co-occurrence~\cite{bar_cortical_2003,bonner_object_2021} and spatial arrangement~\cite{kaiser_transformation_2018}, and the parietal ROI overlaps with PPC/IPS regions sensitive to relative position and configural information~\cite{chafee_representing_2007,ayzenberg_dorsal_2022}. \hlchange{Still, alignment did not consistently improve when GPNs were given explicit spatial information through saccades. One possibility is that, because GPN-B does not receive explicit saccade information, it can only solve the next-glimpse prediction task by learning more location-agnostic scene associations: given one glimpse, what other visual content is likely to occur in the same scene? This kind of location-agnostic association may better match centrally fixated NSD responses than the more specific object-location predictions enabled by GPN-S, even if the latter are useful for next-glimpse prediction. The recurrent models complicate this picture: GPN-R integrates content across glimpses without explicit location information, whereas GPN-RS can use recurrence and saccades to maintain what appeared where across the sequence. This may allow GPN-RS to move from single location-conditioned prediction towards an activity-based memory of multiple glimpse-location bindings, but whether this helps or hurts cortical alignment was backbone-dependent. Thus, the reduction in alignment may reflect limited spatial-arrangement information in NSD responses, a mismatch between GPN and cortical spatial codes, or a difference between location-agnostic scene association and location-specific glimpse prediction.} Disambiguating these accounts is an important next step towards improving models of cortical scene representation, which, despite achieving state-of-the-art alignment, still explain less than half of the variance shared across participants.


In sum, by highlighting the utility of predicting upcoming features during eye movements, our results point towards the next generation of self-supervised models of human cortical scene representation: networks based on the general principle of discovering the structure of the world through temporal, sensorimotor predictions~\cite{elman_finding_1990,oregan_sensorimotor_2001,lecun_path_2022,rao_active_2023}. The success of GPNs originates from taking stock of the signals available to the visual system, rather than following the `computational opportunism' of testing large-scale models engineered towards AI performance, and thereby underlines the importance of cognitive computational neuroscience to emancipate itself, at least to a degree, from AI engineering~\cite{linsley_better_2025}.

\subsection*{Materials \& Methods}

\subsubsection*{Network architecture}

\textit{Glimpse Prediction Networks (GPN)}

Given an ordered sequence of fixations and the associated glimpse {$l_i$, $g_i$}, and a backbone $B$, GPNs predict the backbone’s embedding of the next glimpse, i.e. 
\begin{equation*}
    B(g_{t+1})^{prediction} = \text{GPN}(B(g_t),S\times (l_{t+1}-l_t),R\times ({h_{t-1}})),
\end{equation*}
where S controls if the saccade vector (inter-glimpse displacement) is available, and R controls if the previous hidden state is available. If R is True, GPN contains an LSTM with $3$ layers with hidden-size of $1024$ (implemented with the PyTorch LSTM class)~\cite{paszke_pytorch_2019}. If False, it instead contains the forward pass of the same LSTM with the recurrent components set to zero. The glimpse embedding and saccade vector were transformed to $512$D each, concatenated, and passed on to the LSTM. The LSTM output was projected to a $341$D hidden layer, from which a projection served as the prediction. Layer-norm was used on all layers except the LSTM layers (but it was used on the LSTM output). Dropout was used in all the glimpse representations and in the LSTM. ReLU was used on all non-LSTM layers except the prediction layer.

Glimpse Stitching Networks (GSNs) used the same architecture, but different prediction output sizes, corresponding to their targets—caption embedding for Semantic-GSN (sGSN) and multihot object classification for Classification-GSN (cGSN). The GSNs predicted their scene-level targets given each glimpse embedding input in the sequence, similar to the GPNs predicting the next glimpse embedding per glimpse embedding input. Additionally, for cGSN, which predicts multihot labels of objects in the COCO scenes, no layernorm was used on the prediction layer. The ``stitching into a target representation, using recurrent networks'' process was inspired by previous work~\cite{thompson_zero-shot_2024}. 

Specifically, here, we considered cGSN-R and sGSN-R with SimCLR backbones, that are comparable on architecture and training data to GPN-R, with the RN50-SimCLR backbone, to assess the influence of objective on the alignment of the internal representations of these networks to the human visual cortex.\\

\noindent \textit{GPN architecture search}

The aforementioned architectural configuration was decided by running a hyperparameter sweep over the \#layers ($1$, $3$, $5$) and hidden size ($512$, $1024$, $2048$) of the LSTM, and the input ($d_{in}$) and LSTM dropout ($d_{lstm}$) parameters ($0$, $1$, $0.25$, $0.5$), to optimize the prediction loss for GPN-RS-ILSVRC on the `train'/`val' sets. For all the other GPNs (and for the GSNs), the dropout parameters alone were optimized for lowering prediction loss on the `train'/`val' set. The dropout parameters chosen for each GPN/GSN are listed in Table~\ref{tab:gpn-gsn-vvc}. Final training was performed on the larger `train-515' set and evaluation was carried out on the `test-515' set.

To verify the validity of the trained GSNs, we assessed the alignment of their internal representations to scene representations in ventral visual cortex (VVC). First, as seen in Fig.~\ref{fig:main-neural-results}, we note that the GSNs replicated a previously-reported finding: the caption embedding prediction objective (sGSN) yielded better alignment with VVC than the object class prediction objective (cGSN; mean difference, $\bar{\Delta} = 0.017$, $W = 0$, $p = 0.0078$)~\cite{doerig_high-level_2025}. Additionally, the degree of VVC-alignment of sGSN was comparable (slightly higher, $\bar{\Delta} = 0.02$, $W = 0$, $p = 0.0078$) to the previously-reported alignment based on a different recurrent architecture trained only on full COCO scenes (BLT-MPNet).

\subsubsection*{Network objectives}

\textit{Contrastive prediction loss}

A contrastive prediction loss was used to train the GPNs, as shown in Fig.~\ref{fig:img-net-depiction}A. The positive and negative pairs were sampled from a batch of glimpse sequences from randomly-selected scenes (ordered to form an epoch). Cosine similarity to the target glimpse embedding was maximized, whereas the cosine similarity to the embeddings of other glimpses from the same sequence and glimpses from other sequences was minimized, with equally shared weighting of the average similarities. To avoid minimizing similarity to instances of glimpse refixation, we filtered out negative pairs with cosine similarities higher than $0.999$.  

For sGSN, a cosine similarity loss was used to maximize the alignment to the target caption embedding and minimize the alignment to other scenes' caption embeddings. For cGSN, Binary-Cross-Entropy (BCE) loss with the multihot target vectors was minimized.\\

\noindent \textit{Training details}

The learning rate was reduced on the plateau of the training loss, until it reached $10^{-8}$, when training was stopped. However, prior to this, if validation loss increased more than $1\%$ from the minimum validation loss recorded, training was stopped. The checkpoint with the lowest validation loss was saved.

\subsubsection*{Image datasets}

\textit{COCO scenes \& glimpses}

Scene images were taken from the $2017$ train split of the Common Objects in Context (COCO) dataset. $256\,$px central crops were considered after scaling the smallest side of each image to $256\,$px. DeepGazeIII was used to sample $10$ fixation sequences for each scene (both the order of fixations and the prior distribution of fixations across scenes positively influenced the VVC-alignment results for GPN-R, with the RN50-SimCLR backbone, shown in Fig.~\ref{fig:main-neural-results}; see Fig.~\ref{fig:saccadic-policy-results}). Each sequence contained $7$ fixations, all starting at the center of the images. $91\,$px glimpse crops were extracted centered on these fixations, corresponding to a $3$-degree window (approximately the full-width at half-maximum, FWHM, of retinal cone density)~\cite{legras_distribution_2018}, assuming the image extended $8.4$ degrees of visual field (corresponding to the display parameters from NSD). Semantic embeddings of the captions of the COCO scenes were obtained through MPNet ($768$D)~\cite{doerig_high-level_2025}. Multihot labels signaling the presence of $91$ object classes in the scenes were also obtained from the COCO dataset.

DeepGazeIII and the visual backbones were pretrained and frozen throughout; only the GPN or GSN parameters were optimized in the experiments reported here.

The data was split into three sets: `train' with $\sim 43$k, `val' with $\sim 2$k, `test' with the $\sim 73$k COCO scenes that are part of NSD~\cite{doerig_high-level_2025}. We also created two more splits, `train-515' with $\sim 115$k scenes (pooling from `train' and `test') and `test-515' with the special-515 scenes that each participant saw thrice in NSD. The original, smaller, train/val splits were used for running fast hyperparameter sweeps. The `train-515' split was created to train GPNs/GSNs with the maximum amount of data available that does not overlap with the test set of interest—the special-515 scenes from NSD.\\

\noindent \textit{Backbone feature encoders}

Three encoders were used to extract high-level embeddings of these glimpses: Resnet-50 trained on ILSVRC dataset to classify images (v1)~\cite{he_deep_2015} and one trained with SimCLR~\cite{chen_simple_2020}, and a ViT trained with DINOv2 on LVD-142M (v2, B/14)~\cite{oquab_dinov2_2024}. The AvgPool layer responses of the ResNets ($2048$D), and the output embedding of DINO ($768$D), were used as glimpse embeddings.

Prediction was conducted in high-level embedding space, as opposed to directly on glimpse pixels, as pixel-based prediction would require a much larger generative network and has been shown to rely on features uninformative for high-level perception of objects~\cite{balestriero_contrastive_2022,balestriero_learning_2024}. Pretrained backbones were used, as training the backbone with our procedure did not yield meaningful behavioral and neural alignment results, possibly owing to poor feature learning due to low diversity of the COCO images ($\sim 115$k images), as compared to ILSVRC ($\sim1$M images) or LVD-142M images that the backbones were trained on.

\subsubsection*{Behavioral analysis}

To assess if the GPNs were sensitive to co-occurrence and spatial arrangement of glimpses in natural scenes, and if they could integrate information across glimpses to improve future predictions, we compared the GPN predictions with the relevant target glimpse embeddings. For example, a sensitivity to co-occurrence was indicated by the prediction being better aligned (higher cosine similarity) with embeddings of glimpses from the same scene than from other scenes. These comparisons were conducted within batches of glimpse sequences over the `test-515' scenes. In total, we used 5150 glimpse sequences ($10$ per scene) which were distributed randomly into $25$ batches of $206$ glimpse sequences each.

\subsubsection*{Brain alignment analysis}

\textit{Natural Scenes Dataset}

Neural scene representations were acquired from the Natural Scenes Dataset (NSD). Each of the 8 participants saw up to $10000$ scene images, performing a long-term continuous recognition task. For our analyses comparing GPNs and other models to brain representations, we used the set of $515$ scene images that all $8$ participants saw three times. Each of the $8$ participants' betas (from condition-tied GLMs conducted on the BOLD responses) were transformed to the `fsaverage' space, and betas were acquired from $7$ regions-of-interest (ROIs) from the streams definition, provided in NSD—early, midlateral, midventral, midparietal, ventral, lateral, and parietal. Their definitions, from the NSD manual, can be found in the Supplementary Materials. Betas were averaged across those repeats to get more robust estimates of neural response. Note that the NSD data were collected under the original study's institutional ethics approval and informed-consent procedures; no new human-subject data were collected for this study.\\

\noindent \textit{Alignment based on linear regression} 

To assess the representational alignment between the representations from a brain ROI and a network layer, representational similarity analysis (RSA) was used, which quantifies the alignment between representational geometries i.e., the alignment of principal dimensions of representational variation. Representational dissimilarity matrices (RDMs) were computed from the ROI betas and from the network layer, using pairwise Pearson correlation distances, for the special-515 scenes. The alignment was computed using non-negative linear regression (scikit-learn, positive=True)~\cite{pedregosa_scikit-learn_2011}, as shown in Fig.~\ref{fig:main-neural-results}A. Specifically, the lower triangle of each RDM was flattened, standardized ($\mu=0$, $\sigma=1$), and the linear regression parameters were fitted to minimize the squared error with respect to the flattened lower triangle of the ROI RDM. This choice was made to make the alignment readily amenable to the variance partitioning analysis explained next. The pattern of results looked similar when using Pearson correlation between the brain and model RDMs (Fig.~\ref{fig:pearson-comparison}; which is typically used in RSA)~\cite{kriegeskorte_representational_2008}.\\

\noindent \textit{Variance partitioning analysis} 

Unique variance of an ROI RDM (A) explained by an alternative network RDM (B), compared to variance explained by a GPN RDM (C), is computed as the variance of A explained jointly by B and C minus the variance of A explained by C alone. The variance explained is computed using non-negative linear regression. 

It is important to note that this procedure likely overestimates the unique variance explained by models beyond GPN-R (with the RN50-SimCLR backbone): in multiple regression with correlated predictors, ``suppression'' effects can inflate the total variance explained by the joint model~\cite{conger_revised_1974}. However, in our case, the estimated unique contributions of the alternative models are already very small, so any downward correction would only strengthen our conclusion that they add little beyond the GPN-R.\\

\noindent \textit{Subject shared variance} 

Per ROI, we computed an estimate of the shared variance across subjects as a benchmark of how much variance a single model could capture across subjects in NSD: iteratively, we computed the variance explained (via non-negative linear regression) in the left-out-subject’s RDM by the average RDMs of $N-1$ subjects, and reported the average explained variance. This corresponds to the lower bound of the noise ceiling in typical RSA frameworks.\\

\noindent \textit{Alternative models} 

Brain alignment of the GPNs was compared to a large number of related and state-of-the-art approaches. For each family of networks, we considered openly-available small, medium-sized, and the largest variants. Per ROI, alignment was computed with the RDMs over the special-515 scenes (and also their central $91\,$px crops) per network variant and layer. The best alignment score across variants and their layers was associated with the network family. All the network families and variants tested, and their layers with the best alignment scores to ventral ROI from NSD are summarized in Table~\ref{tab:model-family-overview} and detailed in Tables~\ref{tab:gpn-gsn-vvc}--\ref{tab:vision-language-vvc}.

\subsubsection*{Statistical analyses}

We performed all comparisons on bounded similarity metrics—cosine similarity ($[-1, 1]$) and variance explained ($R^2\in [0,1]$)—using nonparametric tests. Unless stated otherwise, tests were two-sided with $\alpha = 0.05$.\\

\noindent \textit{Paired two-level or one-sample tests} 

We used the Wilcoxon signed-rank test (SciPy)~\cite{virtanen_scipy_2020}. We report the Wilcoxon $W$ statistic (sum of the ranks of the differences above or below zero, whichever is smaller) and the p-value. Zero differences were excluded (SciPy’s default), and exact p-values were used when available; otherwise the large-sample approximation with continuity correction was used.\\

\noindent \textit{$>2$ levels (within-subject)}

We used the Friedman test (SciPy), reporting $\chi^2_{df}$ and the p-value.\\

\noindent \textit{Hierarchical $2\times 2 \times N$ design (recurrence $\times$ saccade $\times$ backbone)} 

For each subject we formed interaction summaries as the average contrast
\begin{equation*}
    \frac{(RS - R) - (S - B)}{2},
\end{equation*}
and tested whether these differed across backbones using the Friedman test. Because the three-way interaction was significant in our data, we did not proceed to simple-effects follow-ups. Main-effect summaries were formed as averages over the other factor, e.g. for recurrence:
\begin{equation*}
    \frac{(R - B) + (RS - S)}{2}
\end{equation*}
and assessed with Wilcoxon signed-rank tests; p-values across main-effect summaries were Holm-adjusted. 

When N = 1, we tested the interaction and both main-effect summaries with Wilcoxon signed-rank tests and applied Holm correction across the three tests.\\

\noindent \textit{Multiple comparisons correction}

When multiple posthoc/independent Wilcoxon/Friedman tests were conducted simultaneously (e.g., several layers in one analysis; data apriori independent), we controlled the false discovery rate with Benjamini–Hochberg (BH-FDR) (statsmodels)~\cite{seabold_statsmodels_2010}. Instead, when correcting a small, pre-specified set of planned contrasts (e.g., main-effect/interaction summaries; data apriori shared and restructured), we controlled the family-wise error rate with Holm adjustment (statsmodels).



\clearpage 

%
\bibliography{main} 

@article{oregan_sensorimotor_2001,
	title = {A {Sensorimotor} {Account} of {Vision} and {Visual} {Consciousness}},
	volume = {24},
	doi = {10.1017/s0140525x01000115},
	number = {5},
	journal = {Behavioral and Brain Sciences},
	author = {O'Regan, J. Kevin and Noë, Alva},
	year = {2001},
	pages = {883--917},
}

@article{doerig_high-level_2025,
	title = {High-level visual representations in the human brain are aligned with large language models},
	volume = {7},
	copyright = {2025 The Author(s)},
	issn = {2522-5839},
	url = {https://www.nature.com/articles/s42256-025-01072-0},
	doi = {10.1038/s42256-025-01072-0},
	language = {en},
	number = {8},
	urldate = {2025-11-24},
	journal = {Nature Machine Intelligence},
	author = {Doerig, Adrien and Kietzmann, Tim C. and Allen, Emily and Wu, Yihan and Naselaris, Thomas and Kay, Kendrick and Charest, Ian},
	month = aug,
	year = {2025},
	note = {Publisher: Nature Publishing Group},
	pages = {1220--1234},
}

@incollection{yarbus_eye_1967,
	address = {Boston, MA},
	title = {Eye {Movements} {During} {Perception} of {Complex} {Objects}},
	isbn = {978-1-4899-5379-7},
	url = {https://doi.org/10.1007/978-1-4899-5379-7_8},
	language = {en},
	urldate = {2025-11-24},
	booktitle = {Eye {Movements} and {Vision}},
	publisher = {Springer US},
	author = {Yarbus, Alfred L.},
	editor = {Yarbus, Alfred L.},
	year = {1967},
	doi = {10.1007/978-1-4899-5379-7_8},
	pages = {171--211},
}

@article{wang_better_2023,
	title = {Better models of human high-level visual cortex emerge from natural language supervision with a large and diverse dataset},
	volume = {5},
	copyright = {2023 The Author(s), under exclusive licence to Springer Nature Limited},
	issn = {2522-5839},
	url = {https://www.nature.com/articles/s42256-023-00753-y},
	doi = {10.1038/s42256-023-00753-y},
	language = {en},
	number = {12},
	urldate = {2025-11-24},
	journal = {Nature Machine Intelligence},
	author = {Wang, Aria Y. and Kay, Kendrick and Naselaris, Thomas and Tarr, Michael J. and Wehbe, Leila},
	month = dec,
	year = {2023},
	note = {Publisher: Nature Publishing Group},
	pages = {1415--1426},
}

@article{virtanen_scipy_2020,
	title = {{SciPy} 1.0: fundamental algorithms for scientific computing in {Python}},
	volume = {17},
	copyright = {2020 The Author(s)},
	issn = {1548-7105},
	shorttitle = {{SciPy} 1.0},
	url = {https://www.nature.com/articles/s41592-019-0686-2},
	doi = {10.1038/s41592-019-0686-2},
	language = {en},
	number = {3},
	urldate = {2025-11-24},
	journal = {Nature Methods},
	author = {Virtanen, Pauli and Gommers, Ralf and Oliphant, Travis E. and Haberland, Matt and Reddy, Tyler and Cournapeau, David and Burovski, Evgeni and Peterson, Pearu and Weckesser, Warren and Bright, Jonathan and van der Walt, Stéfan J. and Brett, Matthew and Wilson, Joshua and Millman, K. Jarrod and Mayorov, Nikolay and Nelson, Andrew R. J. and Jones, Eric and Kern, Robert and Larson, Eric and Carey, C. J. and Polat, İlhan and Feng, Yu and Moore, Eric W. and VanderPlas, Jake and Laxalde, Denis and Perktold, Josef and Cimrman, Robert and Henriksen, Ian and Quintero, E. A. and Harris, Charles R. and Archibald, Anne M. and Ribeiro, Antônio H. and Pedregosa, Fabian and van Mulbregt, Paul},
	month = mar,
	year = {2020},
	note = {Publisher: Nature Publishing Group},
	pages = {261--272},
}

@misc{tschannen_siglip_2025,
	title = {{SigLIP} 2: {Multilingual} {Vision}-{Language} {Encoders} with {Improved} {Semantic} {Understanding}, {Localization}, and {Dense} {Features}},
	shorttitle = {{SigLIP} 2},
	url = {http://arxiv.org/abs/2502.14786},
	doi = {10.48550/arXiv.2502.14786},
	urldate = {2025-11-24},
	publisher = {arXiv},
	author = {Tschannen, Michael and Gritsenko, Alexey and Wang, Xiao and Naeem, Muhammad Ferjad and Alabdulmohsin, Ibrahim and Parthasarathy, Nikhil and Evans, Talfan and Beyer, Lucas and Xia, Ye and Mustafa, Basil and Hénaff, Olivier and Harmsen, Jeremiah and Steiner, Andreas and Zhai, Xiaohua},
	month = feb,
	year = {2025},
	note = {arXiv:2502.14786 [cs]},
}

@article{thompson_zero-shot_2024,
	title = {Zero-shot counting with a dual-stream neural network model},
	volume = {112},
	issn = {0896-6273},
	url = {https://www.sciencedirect.com/science/article/pii/S0896627324007293},
	doi = {10.1016/j.neuron.2024.10.008},
	number = {24},
	urldate = {2025-11-24},
	journal = {Neuron},
	author = {Thompson, Jessica A. F. and Sheahan, Hannah and Dumbalska, Tsvetomira and Sandbrink, Julian D. and Piazza, Manuela and Summerfield, Christopher},
	month = dec,
	year = {2024},
	pages = {4147--4158.e5},
}

@misc{tan_efficientnet_2020,
	title = {{EfficientNet}: {Rethinking} {Model} {Scaling} for {Convolutional} {Neural} {Networks}},
	shorttitle = {{EfficientNet}},
	url = {http://arxiv.org/abs/1905.11946},
	doi = {10.48550/arXiv.1905.11946},
	urldate = {2025-11-24},
	publisher = {arXiv},
	author = {Tan, Mingxing and Le, Quoc V.},
	month = sep,
	year = {2020},
	note = {arXiv:1905.11946 [cs]},
}

@article{summerfield_structure_2020,
	title = {Structure learning and the posterior parietal cortex},
	volume = {184},
	issn = {0301-0082},
	url = {https://www.sciencedirect.com/science/article/pii/S0301008219303351},
	doi = {10.1016/j.pneurobio.2019.101717},
	urldate = {2025-11-24},
	journal = {Progress in Neurobiology},
	author = {Summerfield, Christopher and Luyckx, Fabrice and Sheahan, Hannah},
	month = jan,
	year = {2020},
	pages = {101717},
}

@article{seabold_statsmodels_2010,
	title = {Statsmodels: {Econometric} and {Statistical} {Modeling} with {Python}},
	shorttitle = {Statsmodels},
	url = {https://proceedings.scipy.org/articles/Majora-92bf1922-011},
	doi = {10.25080/Majora-92bf1922-011},
	language = {en},
	urldate = {2025-11-24},
	journal = {SciPy 2010},
	author = {Seabold, Skipper and Perktold, Josef},
	month = may,
	year = {2010},
}

@article{rao_active_2023,
	title = {Active {Predictive} {Coding}: {A} {Unifying} {Neural} {Model} for {Active} {Perception}, {Compositional} {Learning}, and {Hierarchical} {Planning}},
	volume = {36},
	issn = {0899-7667},
	shorttitle = {Active {Predictive} {Coding}},
	url = {https://doi.org/10.1162/neco_a_01627},
	doi = {10.1162/neco_a_01627},
	number = {1},
	urldate = {2025-11-24},
	journal = {Neural Computation},
	author = {Rao, Rajesh P. N. and Gklezakos, Dimitrios C. and Sathish, Vishwas},
	month = dec,
	year = {2023},
	pages = {1--32},
}

@misc{radford_learning_2021,
	title = {Learning {Transferable} {Visual} {Models} {From} {Natural} {Language} {Supervision}},
	url = {http://arxiv.org/abs/2103.00020},
	doi = {10.48550/arXiv.2103.00020},
	urldate = {2025-11-24},
	publisher = {arXiv},
	author = {Radford, Alec and Kim, Jong Wook and Hallacy, Chris and Ramesh, Aditya and Goh, Gabriel and Agarwal, Sandhini and Sastry, Girish and Askell, Amanda and Mishkin, Pamela and Clark, Jack and Krueger, Gretchen and Sutskever, Ilya},
	month = feb,
	year = {2021},
	note = {arXiv:2103.00020 [cs]},
}

@article{peelen_category_2017,
	series = {Special {Issue}: {Concepts}, {Actions} and {Objects}: {Functional} and {Neural} {Perspectives}},
	title = {Category selectivity in human visual cortex: {Beyond} visual object recognition},
	volume = {105},
	issn = {0028-3932},
	shorttitle = {Category selectivity in human visual cortex},
	url = {https://www.sciencedirect.com/science/article/pii/S0028393217301215},
	doi = {10.1016/j.neuropsychologia.2017.03.033},
	urldate = {2025-11-24},
	journal = {Neuropsychologia},
	author = {Peelen, Marius V. and Downing, Paul E.},
	month = oct,
	year = {2017},
	pages = {177--183},
}

@article{pedregosa_scikit-learn_2011,
	title = {Scikit-learn: {Machine} {Learning} in {Python}},
	volume = {12},
	issn = {1533-7928},
	shorttitle = {Scikit-learn},
	url = {http://jmlr.org/papers/v12/pedregosa11a.html},
	number = {85},
	urldate = {2025-11-24},
	journal = {Journal of Machine Learning Research},
	author = {Pedregosa, Fabian and Varoquaux, Gaël and Gramfort, Alexandre and Michel, Vincent and Thirion, Bertrand and Grisel, Olivier and Blondel, Mathieu and Prettenhofer, Peter and Weiss, Ron and Dubourg, Vincent and Vanderplas, Jake and Passos, Alexandre and Cournapeau, David and Brucher, Matthieu and Perrot, Matthieu and Duchesnay, Édouard},
	year = {2011},
	pages = {2825--2830},
}

@misc{paszke_pytorch_2019,
	title = {{PyTorch}: {An} {Imperative} {Style}, {High}-{Performance} {Deep} {Learning} {Library}},
	shorttitle = {{PyTorch}},
	url = {http://arxiv.org/abs/1912.01703},
	doi = {10.48550/arXiv.1912.01703},
	urldate = {2025-11-24},
	publisher = {arXiv},
	author = {Paszke, Adam and Gross, Sam and Massa, Francisco and Lerer, Adam and Bradbury, James and Chanan, Gregory and Killeen, Trevor and Lin, Zeming and Gimelshein, Natalia and Antiga, Luca and Desmaison, Alban and Köpf, Andreas and Yang, Edward and DeVito, Zach and Raison, Martin and Tejani, Alykhan and Chilamkurthy, Sasank and Steiner, Benoit and Fang, Lu and Bai, Junjie and Chintala, Soumith},
	month = dec,
	year = {2019},
	note = {arXiv:1912.01703 [cs]},
}

@misc{oquab_dinov2_2024,
	title = {{DINOv2}: {Learning} {Robust} {Visual} {Features} without {Supervision}},
	shorttitle = {{DINOv2}},
	url = {http://arxiv.org/abs/2304.07193},
	doi = {10.48550/arXiv.2304.07193},
	urldate = {2025-11-24},
	publisher = {arXiv},
	author = {Oquab, Maxime and Darcet, Timothée and Moutakanni, Théo and Vo, Huy and Szafraniec, Marc and Khalidov, Vasil and Fernandez, Pierre and Haziza, Daniel and Massa, Francisco and El-Nouby, Alaaeldin and Assran, Mahmoud and Ballas, Nicolas and Galuba, Wojciech and Howes, Russell and Huang, Po-Yao and Li, Shang-Wen and Misra, Ishan and Rabbat, Michael and Sharma, Vasu and Synnaeve, Gabriel and Xu, Hu and Jegou, Hervé and Mairal, Julien and Labatut, Patrick and Joulin, Armand and Bojanowski, Piotr},
	month = feb,
	year = {2024},
	note = {arXiv:2304.07193 [cs]},
}

@misc{oord_representation_2019,
	title = {Representation {Learning} with {Contrastive} {Predictive} {Coding}},
	url = {http://arxiv.org/abs/1807.03748},
	doi = {10.48550/arXiv.1807.03748},
	urldate = {2025-11-24},
	publisher = {arXiv},
	author = {Oord, Aaron van den and Li, Yazhe and Vinyals, Oriol},
	month = jan,
	year = {2019},
	note = {arXiv:1807.03748 [cs]},
}

@inproceedings{nayman_hardcore-nas_2021,
	title = {{HardCoRe}-{NAS}: {Hard} {Constrained} {diffeRentiable} {Neural} {Architecture} {Search}},
	shorttitle = {{HardCoRe}-{NAS}},
	url = {https://proceedings.mlr.press/v139/nayman21a.html},
	language = {en},
	urldate = {2025-11-24},
	booktitle = {Proceedings of the 38th {International} {Conference} on {Machine} {Learning}},
	publisher = {PMLR},
	author = {Nayman, Niv and Aflalo, Yonathan and Noy, Asaf and Zelnik, Lihi},
	month = jul,
	year = {2021},
	note = {ISSN: 2640-3498},
	pages = {7979--7990},
}

@misc{mikolov_efficient_2013,
	title = {Efficient {Estimation} of {Word} {Representations} in {Vector} {Space}},
	url = {http://arxiv.org/abs/1301.3781},
	doi = {10.48550/arXiv.1301.3781},
	urldate = {2025-11-24},
	publisher = {arXiv},
	author = {Mikolov, Tomas and Chen, Kai and Corrado, Greg and Dean, Jeffrey},
	month = sep,
	year = {2013},
	note = {arXiv:1301.3781 [cs]},
}

@misc{linsley_better_2025,
	title = {Better artificial intelligence does not mean better models of biology},
	url = {http://arxiv.org/abs/2504.16940},
	doi = {10.48550/arXiv.2504.16940},
	urldate = {2025-11-24},
	publisher = {arXiv},
	author = {Linsley, Drew and Feng, Pinyuan and Serre, Thomas},
	month = apr,
	year = {2025},
	note = {arXiv:2504.16940 [q-bio]},
}

@inproceedings{lin_microsoft_2014,
	address = {Cham},
	title = {Microsoft {COCO}: {Common} {Objects} in {Context}},
	isbn = {978-3-319-10602-1},
	shorttitle = {Microsoft {COCO}},
	doi = {10.1007/978-3-319-10602-1_48},
	language = {en},
	booktitle = {Computer {Vision} – {ECCV} 2014},
	publisher = {Springer International Publishing},
	author = {Lin, Tsung-Yi and Maire, Michael and Belongie, Serge and Hays, James and Perona, Pietro and Ramanan, Deva and Dollár, Piotr and Zitnick, C. Lawrence},
	editor = {Fleet, David and Pajdla, Tomas and Schiele, Bernt and Tuytelaars, Tinne},
	year = {2014},
	pages = {740--755},
}

@article{legras_distribution_2018,
	title = {Distribution of cone density, spacing and arrangement in adult healthy retinas with adaptive optics flood illumination},
	volume = {13},
	issn = {1932-6203},
	url = {https://journals.plos.org/plosone/article?id=10.1371/journal.pone.0191141},
	doi = {10.1371/journal.pone.0191141},
	language = {en},
	number = {1},
	urldate = {2025-11-24},
	journal = {PLOS ONE},
	author = {Legras, Richard and Gaudric, Alain and Woog, Kelly},
	month = jan,
	year = {2018},
	note = {Publisher: Public Library of Science},
	pages = {e0191141},
}

@article{lecun_path_2022,
	title = {A {Path} {Towards} {Autonomous} {Machine} {Intelligence} {Version} 0.9.2, 2022-06-27},
	volume = {62},
	language = {en},
	number = {1},
	journal = {OpenReview},
	author = {LeCun, Yann},
	year = {2022},
	pages = {1--62},
}

@article{kummerer_deepgaze_2022,
	title = {{DeepGaze} {III}: {Modeling} free-viewing human scanpaths with deep learning},
	volume = {22},
	issn = {1534-7362},
	shorttitle = {{DeepGaze} {III}},
	url = {https://doi.org/10.1167/jov.22.5.7},
	doi = {10.1167/jov.22.5.7},
	number = {5},
	urldate = {2025-11-24},
	journal = {Journal of Vision},
	author = {Kümmerer, Matthias and Bethge, Matthias and Wallis, Thomas S. A.},
	month = apr,
	year = {2022},
	pages = {7},
}

@article{kriegeskorte_representational_2008,
	title = {Representational similarity analysis - connecting the branches of systems neuroscience},
	volume = {2},
	issn = {1662-5137},
	url = {https://www.frontiersin.org/journals/systems-neuroscience/articles/10.3389/neuro.06.004.2008/full},
	doi = {10.3389/neuro.06.004.2008},
	language = {English},
	urldate = {2025-11-24},
	journal = {Frontiers in Systems Neuroscience},
	author = {Kriegeskorte, Nikolaus and Mur, Marieke and Bandettini, Peter A.},
	month = nov,
	year = {2008},
	note = {Publisher: Frontiers},
}

@article{kaiser_object_2019,
	title = {Object {Vision} in a {Structured} {World}},
	volume = {23},
	issn = {1364-6613},
	url = {https://www.sciencedirect.com/science/article/pii/S1364661319301056},
	doi = {10.1016/j.tics.2019.04.013},
	number = {8},
	urldate = {2025-11-24},
	journal = {Trends in Cognitive Sciences},
	author = {Kaiser, Daniel and Quek, Genevieve L. and Cichy, Radoslaw M. and Peelen, Marius V.},
	month = aug,
	year = {2019},
	pages = {672--685},
}

@article{kaiser_transformation_2018,
	title = {Transformation from independent to integrative coding of multi-object arrangements in human visual cortex},
	volume = {169},
	issn = {1053-8119},
	url = {https://www.sciencedirect.com/science/article/pii/S1053811917310893},
	doi = {10.1016/j.neuroimage.2017.12.065},
	urldate = {2025-11-24},
	journal = {NeuroImage},
	author = {Kaiser, Daniel and Peelen, Marius V.},
	month = apr,
	year = {2018},
	pages = {334--341},
}

@inproceedings{huh_position_2024,
	title = {Position: {The} {Platonic} {Representation} {Hypothesis}},
	shorttitle = {Position},
	url = {https://proceedings.mlr.press/v235/huh24a.html},
	language = {en},
	urldate = {2025-11-24},
	booktitle = {Proceedings of the 41st {International} {Conference} on {Machine} {Learning}},
	publisher = {PMLR},
	author = {Huh, Minyoung and Cheung, Brian and Wang, Tongzhou and Isola, Phillip},
	month = jul,
	year = {2024},
	note = {ISSN: 2640-3498},
	pages = {20617--20642},
}

@misc{he_masked_2021,
	title = {Masked {Autoencoders} {Are} {Scalable} {Vision} {Learners}},
	url = {http://arxiv.org/abs/2111.06377},
	doi = {10.48550/arXiv.2111.06377},
	urldate = {2025-11-24},
	publisher = {arXiv},
	author = {He, Kaiming and Chen, Xinlei and Xie, Saining and Li, Yanghao and Dollár, Piotr and Girshick, Ross},
	month = dec,
	year = {2021},
	note = {arXiv:2111.06377 [cs]},
}

@misc{he_deep_2015,
	title = {Deep {Residual} {Learning} for {Image} {Recognition}},
	url = {http://arxiv.org/abs/1512.03385},
	doi = {10.48550/arXiv.1512.03385},
	urldate = {2025-11-24},
	publisher = {arXiv},
	author = {He, Kaiming and Zhang, Xiangyu and Ren, Shaoqing and Sun, Jian},
	month = dec,
	year = {2015},
	note = {arXiv:1512.03385 [cs]},
}

@article{harris_distributional_1954,
	title = {Distributional structure},
	volume = {10},
	doi = {10.1080/00437956.1954.11659520},
	journal = {Word},
	author = {Harris, Zellig S.},
	year = {1954},
	pages = {146--162},
}

@misc{graham_levit_2021,
	title = {{LeViT}: a {Vision} {Transformer} in {ConvNet}'s {Clothing} for {Faster} {Inference}},
	shorttitle = {{LeViT}},
	url = {http://arxiv.org/abs/2104.01136},
	doi = {10.48550/arXiv.2104.01136},
	urldate = {2025-11-24},
	publisher = {arXiv},
	author = {Graham, Ben and El-Nouby, Alaaeldin and Touvron, Hugo and Stock, Pierre and Joulin, Armand and Jégou, Hervé and Douze, Matthijs},
	month = may,
	year = {2021},
	note = {arXiv:2104.01136 [cs]},
}

@article{frank_bridging_2023,
	title = {Bridging the data gap between children and large language models},
	volume = {27},
	issn = {1364-6613},
	url = {https://www.sciencedirect.com/science/article/pii/S1364661323002036},
	doi = {10.1016/j.tics.2023.08.007},
	number = {11},
	urldate = {2025-11-24},
	journal = {Trends in Cognitive Sciences},
	author = {Frank, Michael C.},
	month = nov,
	year = {2023},
	pages = {990--992},
}

@article{epstein_scene_2019,
	title = {Scene {Perception} in the {Human} {Brain}},
	volume = {5},
	issn = {2374-4642, 2374-4650},
	url = {https://www.annualreviews.org/content/journals/10.1146/annurev-vision-091718-014809},
	doi = {10.1146/annurev-vision-091718-014809},
	language = {en},
	number = {Volume 5, 2019},
	urldate = {2025-11-24},
	journal = {Annual Review of Vision Science},
	author = {Epstein, Russell A. and Baker, Chris I.},
	month = sep,
	year = {2019},
	note = {Publisher: Annual Reviews},
	pages = {373--397},
}

@article{epstein_cortical_1998,
	title = {A cortical representation of the local visual environment},
	volume = {392},
	copyright = {1998 Macmillan Magazines Ltd.},
	issn = {1476-4687},
	url = {https://www.nature.com/articles/33402},
	doi = {10.1038/33402},
	language = {en},
	number = {6676},
	urldate = {2025-11-24},
	journal = {Nature},
	author = {Epstein, Russell and Kanwisher, Nancy},
	month = apr,
	year = {1998},
	note = {Publisher: Nature Publishing Group},
	pages = {598--601},
}

@article{elman_finding_1990,
	title = {Finding {Structure} in {Time}},
	volume = {14},
	issn = {1551-6709},
	url = {https://onlinelibrary.wiley.com/doi/abs/10.1207/s15516709cog1402_1},
	doi = {10.1207/s15516709cog1402_1},
	language = {en},
	number = {2},
	urldate = {2025-11-24},
	journal = {Cognitive Science},
	author = {Elman, Jeffrey L.},
	year = {1990},
	note = {\_eprint: https://onlinelibrary.wiley.com/doi/pdf/10.1207/s15516709cog1402\_1},
	pages = {179--211},
}

@misc{conwell_monkey_2025,
	title = {Monkey {See}, {Model} {Knew}: {Large} {Language} {Models} {Accurately} {Predict} {Visual} {Brain} {Responses} in {Humans} and {Non}-{Human} {Primates}},
	copyright = {© 2025, Posted by Cold Spring Harbor Laboratory. This pre-print is available under a Creative Commons License (Attribution-NonCommercial-NoDerivs 4.0 International), CC BY-NC-ND 4.0, as described at http://creativecommons.org/licenses/by-nc-nd/4.0/},
	shorttitle = {Monkey {See}, {Model} {Knew}},
	url = {https://www.biorxiv.org/content/10.1101/2025.03.05.641284v2},
	doi = {10.1101/2025.03.05.641284},
	language = {en},
	urldate = {2025-11-24},
	publisher = {bioRxiv},
	author = {Conwell, Colin and McMahon, Emalie and Jagadeesh, Akshay and Vinken, Kasper and Sharma, Saloni and Prince, Jacob S. and Alvarez, George A. and Konkle, Talia and Livingstone, Margaret and Isik, Leyla},
	month = apr,
	year = {2025},
	note = {Pages: 2025.03.05.641284
Section: New Results},
}

@article{conwell_large-scale_2024,
	title = {A large-scale examination of inductive biases shaping high-level visual representation in brains and machines},
	volume = {15},
	copyright = {2024 The Author(s)},
	issn = {2041-1723},
	url = {https://www.nature.com/articles/s41467-024-53147-y},
	doi = {10.1038/s41467-024-53147-y},
	language = {en},
	number = {1},
	urldate = {2025-11-24},
	journal = {Nature Communications},
	author = {Conwell, Colin and Prince, Jacob S. and Kay, Kendrick N. and Alvarez, George A. and Konkle, Talia},
	month = oct,
	year = {2024},
	note = {Publisher: Nature Publishing Group},
	pages = {9383},
}

@article{conger_revised_1974,
	title = {A {Revised} {Definition} for {Suppressor} {Variables}: a {Guide} {To} {Their} {Identification} and {Interpretation}},
	volume = {34},
	issn = {0013-1644},
	shorttitle = {A {Revised} {Definition} for {Suppressor} {Variables}},
	url = {https://doi.org/10.1177/001316447403400105},
	doi = {10.1177/001316447403400105},
	language = {EN},
	number = {1},
	urldate = {2025-11-24},
	journal = {Educational and Psychological Measurement},
	author = {Conger, Anthony J.},
	month = apr,
	year = {1974},
	note = {Publisher: SAGE Publications Inc},
	pages = {35--46},
}

@misc{chen_simple_2020,
	title = {A {Simple} {Framework} for {Contrastive} {Learning} of {Visual} {Representations}},
	url = {http://arxiv.org/abs/2002.05709},
	doi = {10.48550/arXiv.2002.05709},
	urldate = {2025-11-24},
	publisher = {arXiv},
	author = {Chen, Ting and Kornblith, Simon and Norouzi, Mohammad and Hinton, Geoffrey},
	month = jul,
	year = {2020},
	note = {arXiv:2002.05709 [cs]},
}

@article{chafee_representing_2007,
	title = {Representing {Spatial} {Relationships} in {Posterior} {Parietal} {Cortex}: {Single} {Neurons} {Code} {Object}-{Referenced} {Position}},
	volume = {17},
	issn = {1047-3211},
	shorttitle = {Representing {Spatial} {Relationships} in {Posterior} {Parietal} {Cortex}},
	url = {https://doi.org/10.1093/cercor/bhm017},
	doi = {10.1093/cercor/bhm017},
	number = {12},
	urldate = {2025-11-24},
	journal = {Cerebral Cortex},
	author = {Chafee, Matthew V. and Averbeck, Bruno B. and Crowe, David A.},
	month = dec,
	year = {2007},
	pages = {2914--2932},
}

@misc{caron_emerging_2021,
	title = {Emerging {Properties} in {Self}-{Supervised} {Vision} {Transformers}},
	url = {http://arxiv.org/abs/2104.14294},
	doi = {10.48550/arXiv.2104.14294},
	urldate = {2025-11-24},
	publisher = {arXiv},
	author = {Caron, Mathilde and Touvron, Hugo and Misra, Ishan and Jégou, Hervé and Mairal, Julien and Bojanowski, Piotr and Joulin, Armand},
	month = may,
	year = {2021},
	note = {arXiv:2104.14294 [cs]},
}

@inproceedings{brown_language_2020,
	title = {Language {Models} are {Few}-{Shot} {Learners}},
	volume = {33},
	url = {https://papers.nips.cc/paper/2020/hash/1457c0d6bfcb4967418bfb8ac142f64a-Abstract.html},
	urldate = {2025-11-24},
	booktitle = {Advances in {Neural} {Information} {Processing} {Systems}},
	publisher = {Curran Associates, Inc.},
	author = {Brown, Tom and Mann, Benjamin and Ryder, Nick and Subbiah, Melanie and Kaplan, Jared D and Dhariwal, Prafulla and Neelakantan, Arvind and Shyam, Pranav and Sastry, Girish and Askell, Amanda and Agarwal, Sandhini and Herbert-Voss, Ariel and Krueger, Gretchen and Henighan, Tom and Child, Rewon and Ramesh, Aditya and Ziegler, Daniel and Wu, Jeffrey and Winter, Clemens and Hesse, Chris and Chen, Mark and Sigler, Eric and Litwin, Mateusz and Gray, Scott and Chess, Benjamin and Clark, Jack and Berner, Christopher and McCandlish, Sam and Radford, Alec and Sutskever, Ilya and Amodei, Dario},
	year = {2020},
	pages = {1877--1901},
}

@article{bonner_object_2021,
	title = {Object representations in the human brain reflect the co-occurrence statistics of vision and language},
	volume = {12},
	copyright = {2021 The Author(s)},
	issn = {2041-1723},
	url = {https://www.nature.com/articles/s41467-021-24368-2},
	doi = {10.1038/s41467-021-24368-2},
	language = {en},
	number = {1},
	urldate = {2025-11-24},
	journal = {Nature Communications},
	author = {Bonner, Michael F. and Epstein, Russell A.},
	month = jul,
	year = {2021},
	note = {Publisher: Nature Publishing Group},
	pages = {4081},
}

@article{bar_cortical_2003,
	title = {Cortical {Analysis} of {Visual} {Context}},
	volume = {38},
	issn = {0896-6273},
	url = {https://www.cell.com/neuron/abstract/S0896-6273(03)00167-3},
	doi = {10.1016/S0896-6273(03)00167-3},
	language = {English},
	number = {2},
	urldate = {2025-11-24},
	journal = {Neuron},
	author = {Bar, Moshe and Aminoff, Elissa},
	month = apr,
	year = {2003},
	pmid = {12718867},
	note = {Publisher: Elsevier},
	pages = {347--358},
}

@misc{balestriero_learning_2024,
	title = {Learning by {Reconstruction} {Produces} {Uninformative} {Features} {For} {Perception}},
	url = {https://arxiv.org/abs/2402.11337v1},
	language = {en},
	urldate = {2025-11-24},
	journal = {arXiv.org},
	author = {Balestriero, Randall and LeCun, Yann},
	month = feb,
	year = {2024},
}

@article{balestriero_contrastive_2022,
	title = {Contrastive and {Non}-{Contrastive} {Self}-{Supervised} {Learning} {Recover} {Global} and {Local} {Spectral} {Embedding} {Methods}},
	volume = {35},
	url = {https://proceedings.neurips.cc/paper_files/paper/2022/hash/aa56c74513a5e35768a11f4e82dd7ffb-Abstract-Conference.html},
	language = {en},
	urldate = {2025-11-24},
	journal = {Advances in Neural Information Processing Systems},
	author = {Balestriero, Randall and LeCun, Yann},
	month = dec,
	year = {2022},
	pages = {26671--26685},
}

@inproceedings{bakhtiari_functional_2021,
	title = {The functional specialization of visual cortex emerges from training parallel pathways with self-supervised predictive learning},
	volume = {34},
	url = {https://proceedings.neurips.cc/paper/2021/hash/d384dec9f5f7a64a36b5c8f03b8a6d92-Abstract.html},
	urldate = {2025-11-24},
	booktitle = {Advances in {Neural} {Information} {Processing} {Systems}},
	publisher = {Curran Associates, Inc.},
	author = {Bakhtiari, Shahab and Mineault, Patrick and Lillicrap, Timothy and Pack, Christopher and Richards, Blake},
	year = {2021},
	pages = {25164--25178},
}

@article{ayzenberg_dorsal_2022,
	title = {The {Dorsal} {Visual} {Pathway} {Represents} {Object}-{Centered} {Spatial} {Relations} for {Object} {Recognition}},
	volume = {42},
	issn = {0270-6474, 1529-2401},
	url = {https://www.jneurosci.org/lookup/doi/10.1523/JNEUROSCI.2257-21.2022},
	doi = {10.1523/JNEUROSCI.2257-21.2022},
	language = {en},
	number = {23},
	urldate = {2025-11-24},
	journal = {The Journal of Neuroscience},
	author = {Ayzenberg, Vladislav and Behrmann, Marlene},
	month = jun,
	year = {2022},
	pages = {4693--4710},
}

@misc{assran_self-supervised_2023,
	title = {Self-{Supervised} {Learning} from {Images} with a {Joint}-{Embedding} {Predictive} {Architecture}},
	copyright = {arXiv.org perpetual, non-exclusive license},
	url = {https://arxiv.org/abs/2301.08243},
	doi = {10.48550/ARXIV.2301.08243},
	urldate = {2025-11-24},
	publisher = {arXiv},
	author = {Assran, Mahmoud and Duval, Quentin and Misra, Ishan and Bojanowski, Piotr and Vincent, Pascal and Rabbat, Michael and LeCun, Yann and Ballas, Nicolas},
	year = {2023},
	note = {Version Number: 3},
}

@article{yamins_using_2016,
	title = {Using goal-driven deep learning models to understand sensory cortex},
	volume = {19},
	issn = {1097-6256, 1546-1726},
	url = {https://www.nature.com/articles/nn.4244},
	doi = {10.1038/nn.4244},
	language = {en},
	number = {3},
	urldate = {2025-11-24},
	journal = {Nature Neuroscience},
	author = {Yamins, Daniel L K and DiCarlo, James J},
	month = mar,
	year = {2016},
	pages = {356--365},
}

@article{doerig_neuroconnectionist_2023,
	title = {The neuroconnectionist research programme},
	volume = {24},
	issn = {1471-003X, 1471-0048},
	url = {https://www.nature.com/articles/s41583-023-00705-w},
	doi = {10.1038/s41583-023-00705-w},
	language = {en},
	number = {7},
	urldate = {2025-11-24},
	journal = {Nature Reviews Neuroscience},
	author = {Doerig, Adrien and Sommers, Rowan P. and Seeliger, Katja and Richards, Blake and Ismael, Jenann and Lindsay, Grace W. and Kording, Konrad P. and Konkle, Talia and Van Gerven, Marcel A. J. and Kriegeskorte, Nikolaus and Kietzmann, Tim C.},
	month = jul,
	year = {2023},
	pages = {431--450},
}

@article{richards_deep_2019,
	title = {A deep learning framework for neuroscience},
	volume = {22},
	issn = {1097-6256, 1546-1726},
	url = {https://www.nature.com/articles/s41593-019-0520-2},
	doi = {10.1038/s41593-019-0520-2},
	language = {en},
	number = {11},
	urldate = {2025-11-24},
	journal = {Nature Neuroscience},
	author = {Richards, Blake A. and Lillicrap, Timothy P. and Beaudoin, Philippe and Bengio, Yoshua and Bogacz, Rafal and Christensen, Amelia and Clopath, Claudia and Costa, Rui Ponte and De Berker, Archy and Ganguli, Surya and Gillon, Colleen J. and Hafner, Danijar and Kepecs, Adam and Kriegeskorte, Nikolaus and Latham, Peter and Lindsay, Grace W. and Miller, Kenneth D. and Naud, Richard and Pack, Christopher C. and Poirazi, Panayiota and Roelfsema, Pieter and Sacramento, João and Saxe, Andrew and Scellier, Benjamin and Schapiro, Anna C. and Senn, Walter and Wayne, Greg and Yamins, Daniel and Zenke, Friedemann and Zylberberg, Joel and Therien, Denis and Kording, Konrad P.},
	month = nov,
	year = {2019},
	pages = {1761--1770},
}

@article{allen_massive_2022,
	title = {A massive {7T} {fMRI} dataset to bridge cognitive neuroscience and artificial intelligence},
	volume = {25},
	issn = {1097-6256, 1546-1726},
	url = {https://www.nature.com/articles/s41593-021-00962-x},
	doi = {10.1038/s41593-021-00962-x},
	language = {en},
	number = {1},
	urldate = {2025-11-24},
	journal = {Nature Neuroscience},
	author = {Allen, Emily J. and St-Yves, Ghislain and Wu, Yihan and Breedlove, Jesse L. and Prince, Jacob S. and Dowdle, Logan T. and Nau, Matthias and Caron, Brad and Pestilli, Franco and Charest, Ian and Hutchinson, J. Benjamin and Naselaris, Thomas and Kay, Kendrick},
	month = jan,
	year = {2022},
	pages = {116--126},
}

@article{lu2026adopting,
  title={Adopting a human developmental visual diet yields robust and shape-based AI vision},
  author={Lu, Zejin and Thorat, Sushrut and Cichy, Radoslaw M and Kietzmann, Tim C},
  journal={Nature Machine Intelligence},
  pages={1--14},
  year={2026},
  publisher={Nature Publishing Group UK London}
}

@article{kubilius2018cornet,
  title={Cornet: Modeling the neural mechanisms of core object recognition},
  author={Kubilius, Jonas and Schrimpf, Martin and Nayebi, Aran and Bear, Daniel and Yamins, Daniel LK and DiCarlo, James J},
  journal={BioRxiv},
  pages={408385},
  year={2018},
  publisher={Cold Spring Harbor Laboratory}
}

@article{smolensky1990tensor,
  title={Tensor product variable binding and the representation of symbolic structures in connectionist systems},
  author={Smolensky, Paul},
  journal={Artificial intelligence},
  volume={46},
  number={1-2},
  pages={159--216},
  year={1990},
  publisher={Elsevier}
}

@inproceedings{liu2022convnet,
  title={A convnet for the 2020s},
  author={Liu, Zhuang and Mao, Hanzi and Wu, Chao-Yuan and Feichtenhofer, Christoph and Darrell, Trevor and Xie, Saining},
  booktitle={Proceedings of the IEEE/CVF conference on computer vision and pattern recognition},
  pages={11976--11986},
  year={2022}
}

@article{dosovitskiy2020image,
  title={An image is worth 16x16 words: Transformers for image recognition at scale},
  author={Dosovitskiy, Alexey and Beyer, Lucas and Kolesnikov, Alexander and Weissenborn, Dirk and Zhai, Xiaohua and Unterthiner, Thomas and Dehghani, Mostafa and Minderer, Matthias and Heigold, Georg and Gelly, Sylvain and others},
  journal={arXiv preprint arXiv:2010.11929},
  year={2020}
}

@inproceedings{tu2022maxvit,
  title={Maxvit: Multi-axis vision transformer},
  author={Tu, Zhengzhong and Talebi, Hossein and Zhang, Han and Yang, Feng and Milanfar, Peyman and Bovik, Alan and Li, Yinxiao},
  booktitle={European conference on computer vision},
  pages={459--479},
  year={2022},
  organization={Springer}
}

@inproceedings{liu2021swin,
  title={Swin transformer: Hierarchical vision transformer using shifted windows},
  author={Liu, Ze and Lin, Yutong and Cao, Yue and Hu, Han and Wei, Yixuan and Zhang, Zheng and Lin, Stephen and Guo, Baining},
  booktitle={Proceedings of the IEEE/CVF international conference on computer vision},
  pages={10012--10022},
  year={2021}
}
\bibliographystyle{sciencemag}

%
%
%
%
%
%


\section*{Acknowledgments}
We would like to thank Victoria Bosch for the discussion that kickstarted the project, and all members of Kietzmann lab for the discussions that enriched the proceedings. 

\paragraph*{Funding:}
European Research Council's (ERC) Starting grant \#101039524 ``TIME'' (AD, CA, AK, TCK); Schweizerischer Nationalfonds (SNF) grant \#203018 (AD). 

\paragraph*{Author contributions:}

ST, AD, AK, CA, TCK designed research; ST, AD, AK contributed analytic tools; ST performed research; ST analyzed data; ST, AD, AK, CA, TCK wrote the paper.

\paragraph*{Competing interests:}
The authors declare they have no competing interests.

\paragraph*{Data and materials availability:}
Open-source datasets and comparison networks were used. All the data required to run the analyses reported in this paper are available at: \url{https://doi.org/10.5281/zenodo.20667385}. The scripts to train GPNs and GSNs, and to evaluate them and other networks, are also available at: \url{https://github.com/KietzmannLab/GPN}. No new biological or physical materials were generated in this study. New computational materials generated for this study, including trained model outputs and derived analysis files, are described in the Materials and Methods and are available through the repository and data archive listed above.


\subsection*{Supplementary materials}
Figs. S1 to S11\\
Table S1 to S7


\newpage


\renewcommand{\thefigure}{S\arabic{figure}}
\renewcommand{\thetable}{S\arabic{table}}
\renewcommand{\theequation}{S\arabic{equation}}
\renewcommand{\thepage}{S\arabic{page}}
\setcounter{figure}{0}
\setcounter{table}{0}
\setcounter{equation}{0}
\setcounter{page}{1} 


\begin{center}
\section*{Supplementary Materials for\\ \scititle}

Sushrut Thorat$^\ast$,
Adrien Doerig,
Alexander Kroner,
Carmen Amme,
Tim C. Kietzmann$^\ast$\\
\small$^\ast$Corresponding authors. Emails: sthorat$@$uos.de, tkietzma$@$uos.de
\end{center}

\subsubsection*{This PDF file includes:}
Figs. S1 to S11\\
Table S1 to S7

\newpage



\begin{figure} 
	\centering
	\includegraphics[width=0.3\textwidth]{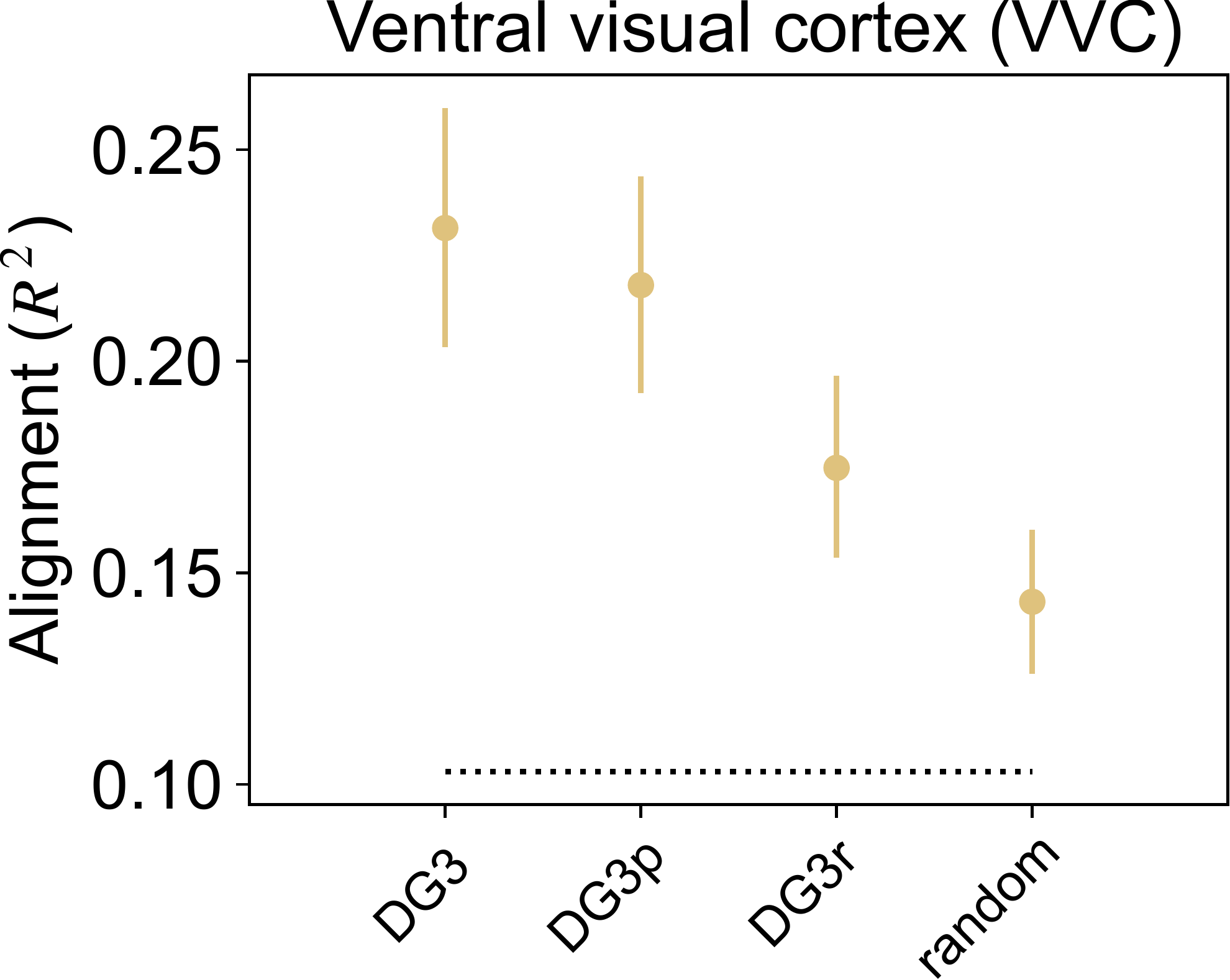} 

	\caption{\textbf{VVC-alignment of variants of GPN-R, with the RN50-SimCLR backbone, trained with differing glimpse sequences.} For GPN-R, glimpses were sampled using DeepGazeIII (DG3). To assess the importance of DG3 fixations in generating VVC-aligned scene representations, seen in Fig.~\ref{fig:main-neural-results}, we considered three more conditions. (i) DG3p: within each fixation sequence, the fixation order was permuted (barring glimpse 0), (ii) DG3r: the fixation locations per sequence were randomly sampled from all possible fixation locations across the scenes (barring glimpse 0), capturing the prior distribution of DG3 fixations (center-biased), and (iii) random: the fixation locations were sampled uniformly-randomly from the scenes (barring glimpse 0). Six pairwise comparisons revealed that (i) the prior center-biased fixation distribution of DG3 was important to acquire VVC-aligned scene representations (DG3r - random; mean difference, $\bar{\Delta}$ = 0.031, W = 0, FDR-adjusted p = 0.0078); (ii) scene-conditioned fixation distributions increased the alignment (DG3p - DG3r; $\bar{\Delta}$ = 0.043, W = 0, FDR-adjusted p = 0.0078); and (iii) the sequence of scene-conditioned fixations further increased alignment (DG3 - DG3p; $\bar{\Delta}$ = 0.013, W = 0, FDR-adjusted p = 0.0078). Notably, the variant with random glimpses still improved alignment with VVC over the max-alignment of input glimpse embeddings (for the central glimpse; dotted line; $\bar{\Delta}$ = 0.04, W = 0, p = 0.0078). These results suggest the glimpse prediction objective improved VVC-alignment, even when random glimpses were provided. However, the use of DeepGazeIII fixation sequences was critical to the high VVC-alignment of emergent scene representations in GPN-R.}
	\label{fig:saccadic-policy-results}
\end{figure}

\begin{figure} 
	\centering
	\includegraphics[width=0.8\textwidth]{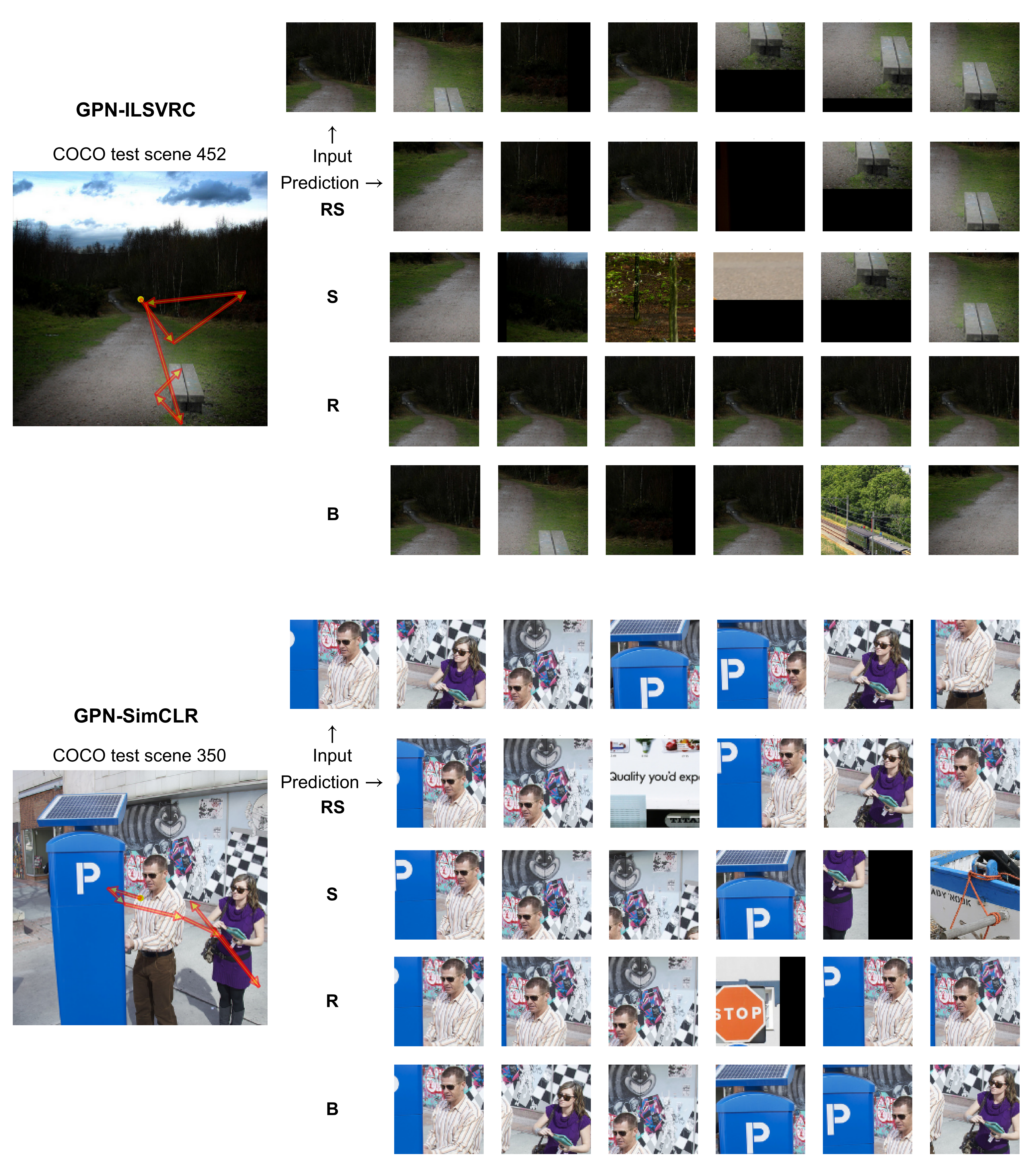} 

\caption{\textbf{Qualitative assessment of the behavior of GPNs across variants and backbones} (see Fig.~\ref{fig:beh-analysis}A for reference).}
	\label{fig:backbone-2-examples}
\end{figure}

\begin{figure} 
	\centering
	\includegraphics[width=0.8\textwidth]{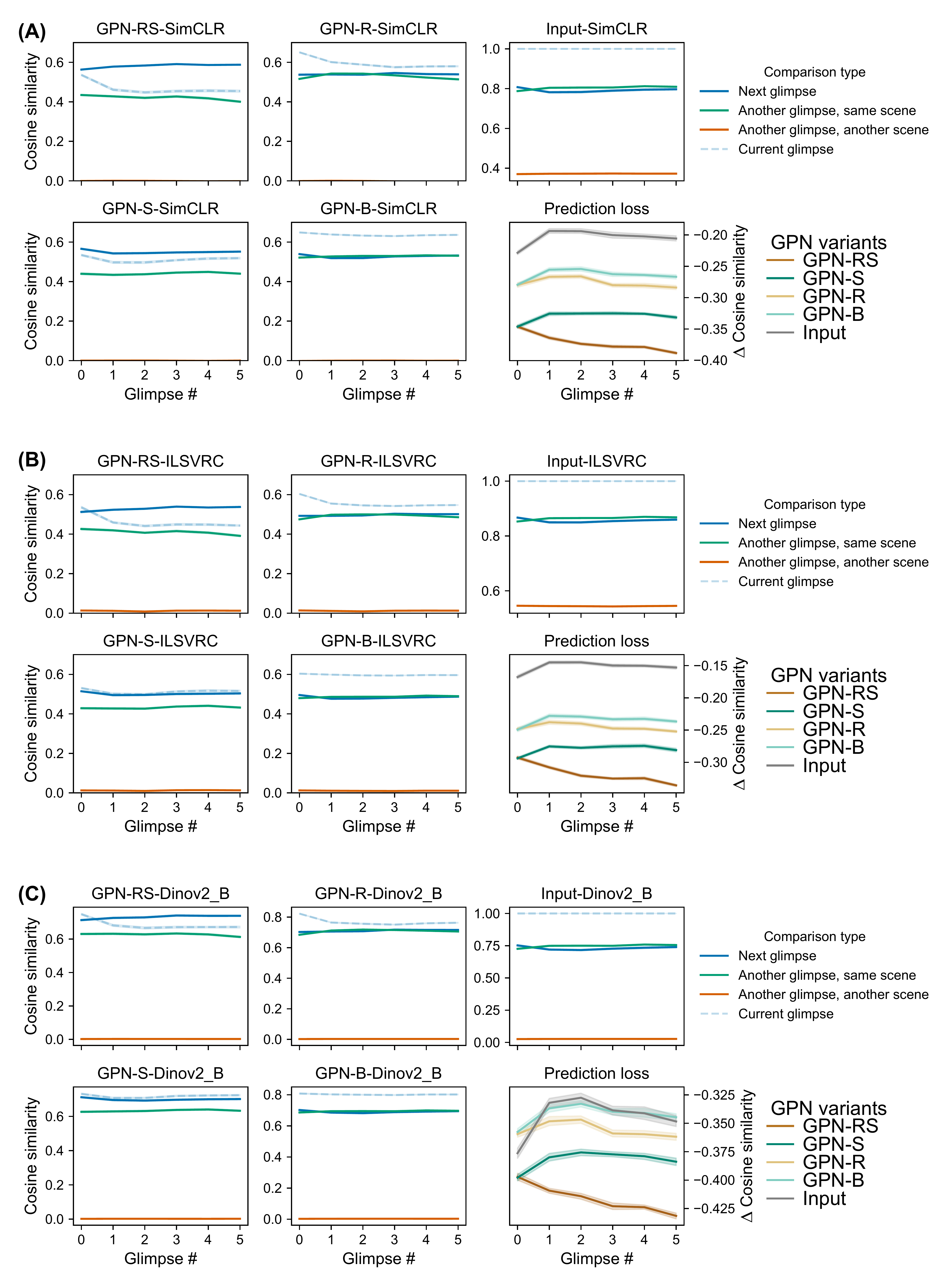} 

\caption{\textbf{Quantitative assessment of the behavior of GPNs across variants} (see Fig.~\ref{fig:beh-analysis}B for reference) for, \textbf{(A)} the SimCLR backbone, \textbf{(B)} the ILSVRC backbone, and \textbf{(C)} the DINOv2-B backbone.}
	\label{fig:backbone-2-behavior}
\end{figure}

\begin{figure} 
	\centering
	\includegraphics[width=0.8\textwidth]{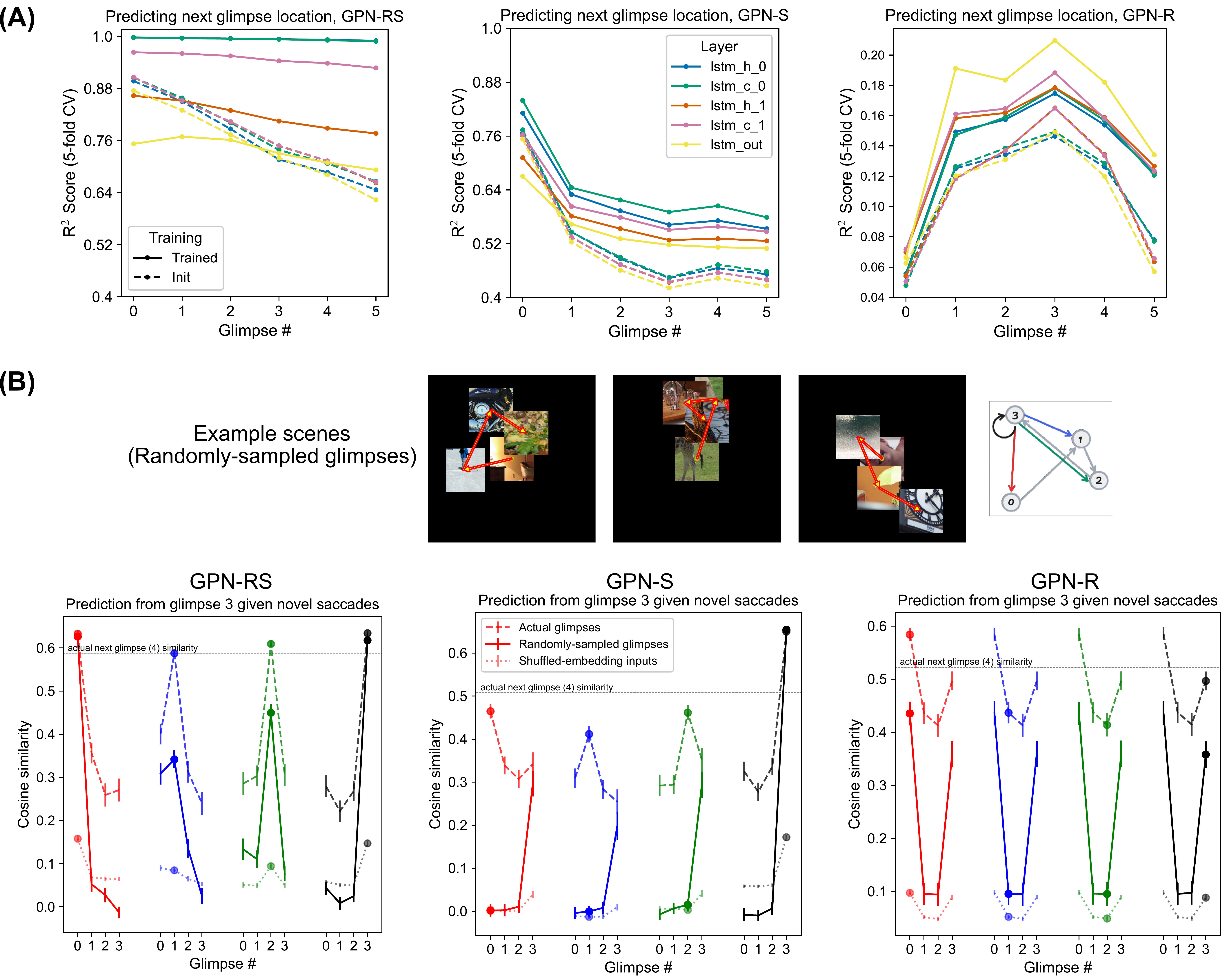} 

\caption{\textbf{Behavioral assessment of glimpse integration (SimCLR backbone).} As shown in Fig.~\ref{fig:beh-analysis}B-C, GPN-RS leveraged spatial arrangement and integrated across glimpses to improve future-glimpse prediction. A plausible mechanism is the path integration of saccades that anchors glimpse embeddings to absolute locations, i.e. tracking what is where. (A) We fit 5-fold cross-validated linear regressions to predict absolute position (x,y; $R^2$ averaged across axes) from model representations across layers and glimpse numbers in the sequences. In GPN-RS, variance explained in the first LSTM layer stayed near 1 as glimpses accrued, consistent with path integration. This effect was absent in the initialized GPN-RS (`Init') and in GPNs-R/S after training, indicating that recurrence and saccades are required to learn path integration. (B) Path integration implies that refixations should cue the correct content at that location. We tested three conditions over 1,000 fixation sequences (first four fixations) from the test-515 set (example scenes are depicted), simulating refixations from glimpse \#3 toward each of the four locations (setup inset shown on the right): (i) Actual glimpses (natural co-occurrence/arrangement), (ii) Randomly-sampled glimpses (glimpse embeddings across scenes randomly assigned to fixations to violate co-occurrence and arrangement), and (iii) Shuffled-embedding inputs (embedding dimensions independently shuffled per glimpse to destroy natural glimpse features). GPN-RS retrieved the correct target embedding under all three conditions; GPN-S succeeded only for `Actual glimpses', as expected without recurrent integration. GPN-R could not succeed for any condition, as expected without saccadic inputs. Together, these results indicate that GPN-RS learned a general embedding-location binding mechanism~\cite{smolensky1990tensor}. Error bars in (B) show 95\% CIs.}
	\label{fig:gpn-rs-binding}
\end{figure}

\begin{figure} 
	\centering
	\includegraphics[width=0.8\textwidth]{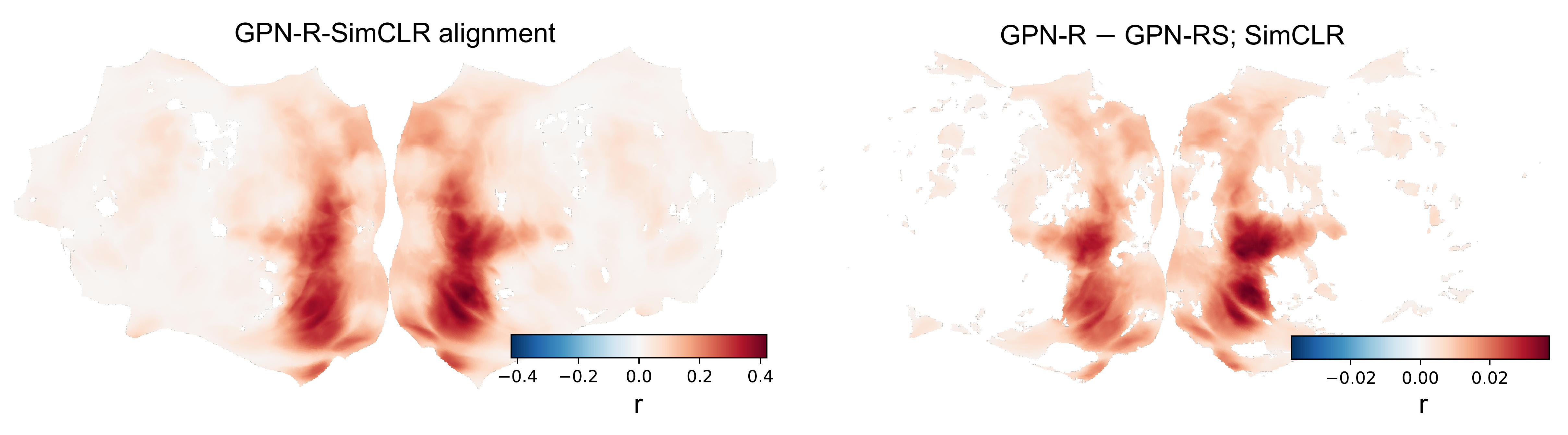} 

\caption{\textbf{ Searchlight analysis to compare the alignment of GPN-R vs GPN-RS (RN50-SimCLR backbone) across the brain.} Searchlight spheres with a radius of 6 voxels, i.e. covering ~900 voxels, were used. Pearson correlation was used to quantify alignment. No spheres revealed a higher alignment with GPN-RS than with GPN-R. Correlations for spheres that were significant ($p < 0.05$; FDR-adjusted across the brain) are shown.}
	\label{fig:searchlight-analysis}
\end{figure}

\begin{figure} 
	\centering
	\includegraphics[width=0.8\textwidth]{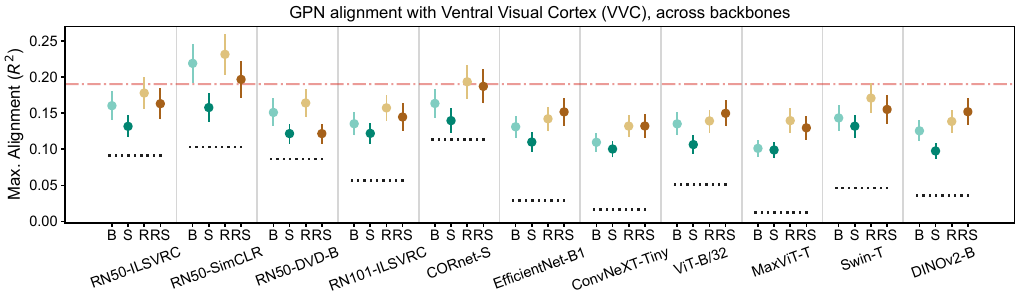} 

\caption{\textbf{VVC-alignment of GPN variants trained over different backbones.} In Fig.~\ref{fig:main-neural-results}B, we saw that with the ResNet backbones, GPN-RS had a lower VVC-alignment than GPN-R, whereas with the DINO backbone, GPN-RS had a higher VVC-alignment than GPN-R. To shed further light on this RS vs R effect as a function of backbone, we trained GPNs with eight more backbones. This included two more ResNet backbones, one trained on ILSVRC (RN101-ILSVRC) and the other trained on an infant-like curriculum using Ecoset images (RN50-DVD-B) which yields a shape-biased model~\cite{lu2026adopting}. It further included a recurrent convolutional neural network (CNN; CORnet-S), which is one of the best models of ventral stream object representations~\cite{kubilius2018cornet}, two more CNNs (EfficientNet-B1 and ConvNeXT-Tiny~\cite{liu2022convnet}), and three more vision transformers (ViT-B/32, MaxViT-T, and Swin-T~\cite{dosovitskiy2020image,tu2022maxvit,liu2021swin}). The degree of VVC-alignment depended on an interaction of the inclusion of recurrence and saccades, and the backbone ($2\times 2\times 11$ repeated-measures analysis; $3$-way interaction: $\chi^2_{10} = 76.1$, $p = 3\times 10^{-12}$). The inclusion of recurrence on top of GPN-B or GPN-S increased VVC-alignment across all backbones ($\bar{\Delta} > 0.01$, $W = 0$, FDR-adjusted $p = 0.0086$), except for GPN-S with the DVD-B backbone where there was no difference ($\bar{\Delta} = 0.0$, $W = 0$, FDR-adjusted $p = 1.0$). 
In contrast, for all backbones, except MaxViT-T, the inclusion of saccades on top of GPN-B reduced VVC-alignment ($\bar{\Delta} < -0.009$, $W = 0$, FDR-adjusted $p = 0.0086$; for MaxViT-T, $\bar{\Delta} = -0.002$, $W = 7$, FDR-adjusted $p = 0.15$). 
However, the inclusion of saccades on top of GPN-R had a mixed effect: all the ResNet, CorNet-S, MaxViT-T, and Swin-T backbones showed reduced VVC-alignment ($\bar{\Delta} < -0.006$, $W = 0$, FDR-adjusted $p = 0.0086$), whereas the EfficientNet-B1, ViT-B/32, and DINOv2-B/14 backbones showed increased VVC-alignment ($\bar{\Delta} > 0.009$, $W = 0$, FDR-adjusted $p = 0.0086$), and the ConvNeXT-Tiny backbone showed no difference ($\bar{\Delta} = 6\times 10^{-5}$, $W = 16$, FDR-adjusted $p = 0.84$). These results suggest that the reduction in VVC-alignment with saccade inclusion over GPN-R was not specific to the ResNet backbones, and we could not identify any clear pattern.}
	\label{fig:gpn-backbones-sweep}
\end{figure}

\begin{figure} 
	\centering
	\includegraphics[width=0.8\textwidth]{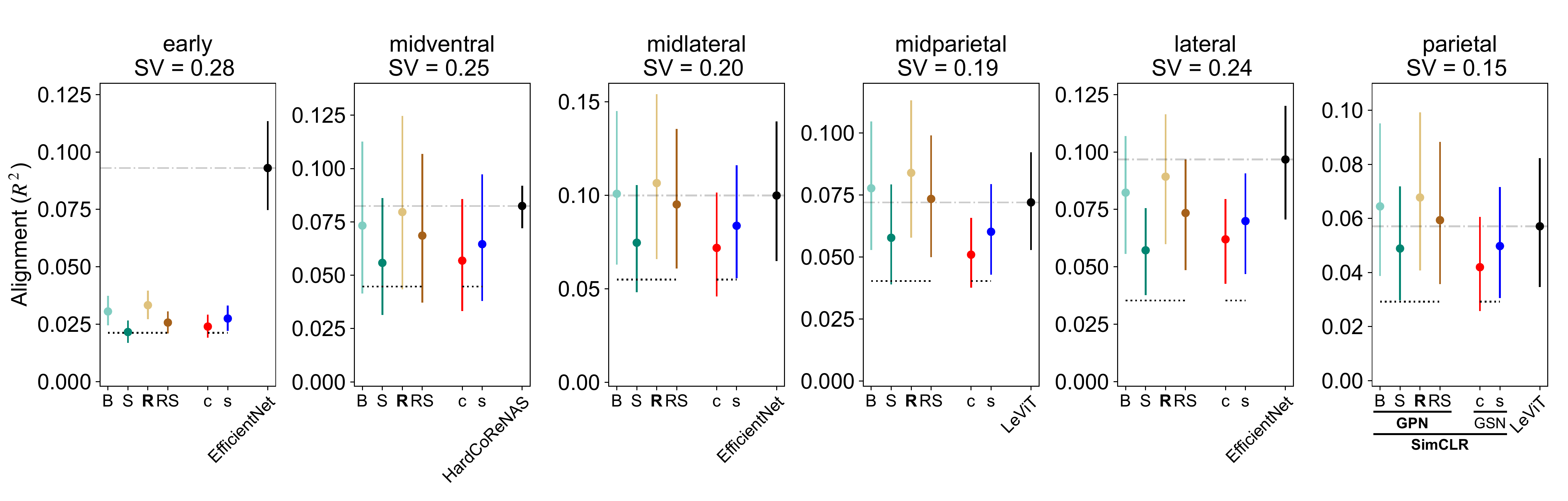} 

\caption{\textbf{Alignment of GPNs with SimCLR backbone and comparison networks across the streams
ROIs} (see Fig.~\ref{fig:main-neural-results} for reference). Per ROI, the alignment of the GSNs and the best-aligned alternative
network for that ROI are shown.}
	\label{fig:comparison-allROIs}
\end{figure}

\begin{figure} 
	\centering
	\includegraphics[width=0.8\textwidth]{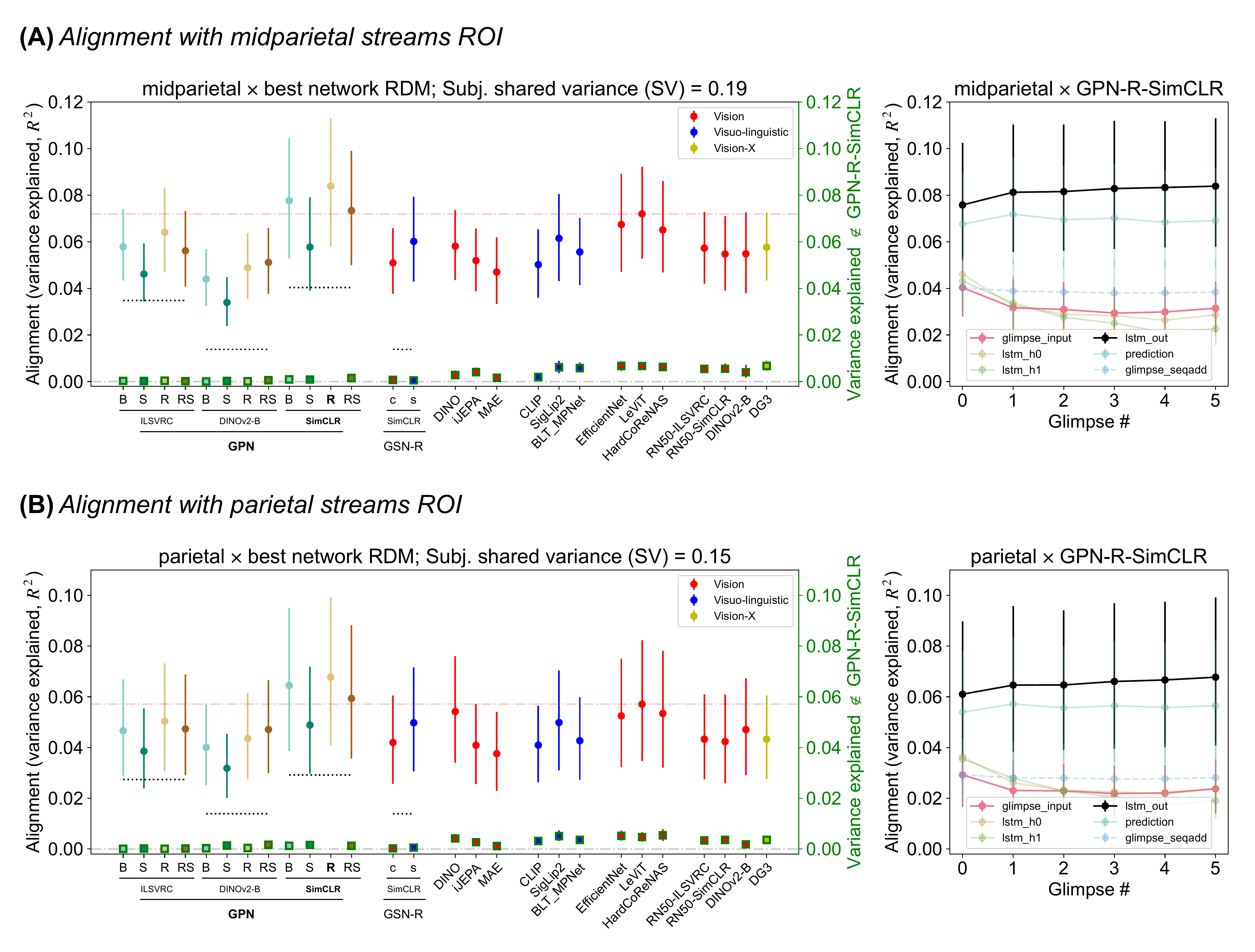} 

\caption{\textbf{Alignment of the GPN variants and comparison networks for the (A) midparietal, and (B) parietal streams ROIs} (see Fig.~\ref{fig:main-neural-results}B for reference).}
	\label{fig:comparison-midparietal-parietal}
\end{figure}

\begin{figure} 
	\centering
	\includegraphics[width=0.8\textwidth]{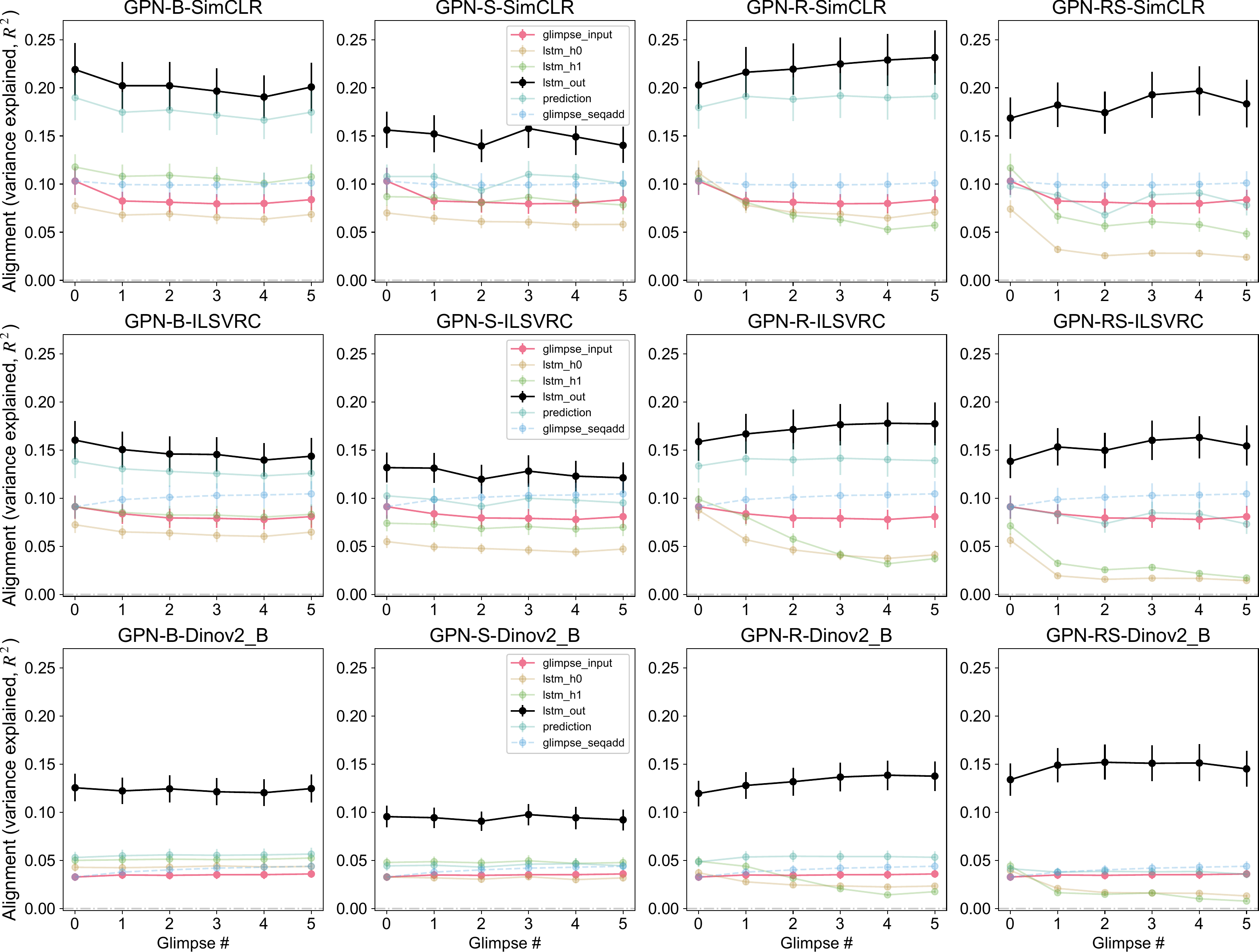} 

\caption{\textbf{VVC-alignment of GPN variants and backbones across layer and glimpse numbers in the sequences}. See Fig.~\ref{fig:gpn-r-simclr-internals}B, for reference.}
	\label{fig:gpn-internals-all}
\end{figure}

\begin{figure} 
	\centering
	\includegraphics[width=0.8\textwidth]{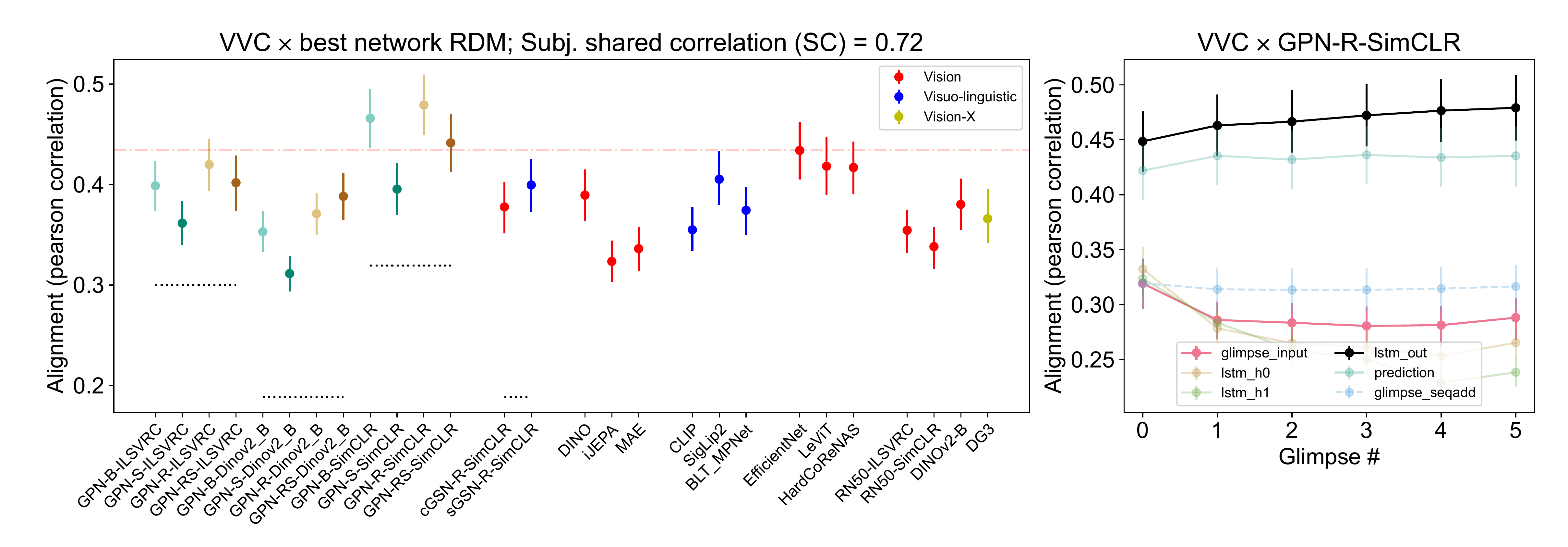} 

\caption{\textbf{VVC-alignment of GPNs and comparison networks, using Pearson correlation, instead of non-negative linear regression's $R^2$, as the measure} (see Fig.~\ref{fig:main-neural-results}B for reference).}
	\label{fig:pearson-comparison}
\end{figure}

\begin{figure} 
	\centering
	\includegraphics[width=0.8\textwidth]{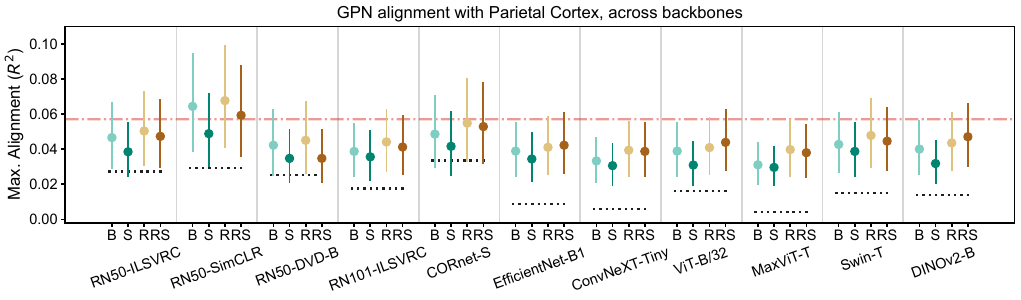} 

\caption{\textbf{Parietal Cortex-alignment of GPN variants trained over different backbones.} See Fig.~\ref{fig:gpn-backbones-sweep}, for methodological details.}
	\label{fig:gpn-backbones-sweep-parietal}
\end{figure}


\begin{table}
	\centering
	\caption{\textbf{Overview of network families tested for VVC alignment.}
		The detailed model-level alignment results are split across Tables~\ref{tab:gpn-gsn-vvc}--\ref{tab:vision-language-vvc}. Each detailed table reports the best-aligned input/layer and the average VVC alignment.}
	\label{tab:model-family-overview}

    \footnotesize
	\begin{tabular}{p{0.24\textwidth}p{0.38\textwidth}p{0.22\textwidth}c}
		\\
		\hline
		{\raggedright Model family\par} & {\raggedright Models included\par} & {\raggedright Purpose\par} & Details\\
		\hline
		{\raggedright GPNs and matched controls\par} & {\raggedright GPN variants across RN50-ILSVRC, RN50-SimCLR, and DINOv2-B; cGSN-R-SimCLR; sGSN-R-SimCLR\par} & {\raggedright Proposed models and matched objective controls\par} & Table~\ref{tab:gpn-gsn-vvc}\\
		\hline
		{\raggedright DINO-family SSL vision models\par} & {\raggedright DINOv2, DINO-WebSSL, and DINOv3 variants\par} & {\raggedright Strong vision-only self-supervised alternatives\par} & Table~\ref{tab:vision-ssl-dino-vvc}\\
		\hline
		{\raggedright iJEPA and MAE SSL vision models\par} & {\raggedright iJEPA and MAE variants, including WebSSL MAE models\par} & {\raggedright Additional predictive and reconstruction-based SSL alternatives\par} & Table~\ref{tab:vision-ssl-ijepa-mae-vvc}\\
		\hline
		{\raggedright Supervised and architecture-optimized vision models\par} & {\raggedright EfficientNet, LeViT, and HardCoreNAS variants\par} & {\raggedright State-of-the-art visual model families\par} & Table~\ref{tab:vision-sota-vvc}\\
		\hline
		{\raggedright Visual baselines\par} & {\raggedright RN50-ILSVRC, RN50-SimCLR, and DeepGazeIII\par} & {\raggedright Backbone and fixation-related baselines\par} & Table~\ref{tab:visual-baselines-vvc}\\
		\hline
		{\raggedright Vision-language and semantic baselines\par} & {\raggedright CLIP, SigLIP2, and BLT-MPNet\par} & {\raggedright Language- or semantic-supervision baselines\par} & Table~\ref{tab:vision-language-vvc}\\
		\hline
	\end{tabular}
\end{table}

\begin{table}
	\centering
	\caption{\textbf{GPNs and matched Glimpse Stitching Networks tested for VVC alignment.}
		For GPNs and GSNs, $(d_{in}, d_{lstm})$ reports the input and LSTM dropout values. The input/layer column reports the input and layer with the highest average VVC alignment.}
	\label{tab:gpn-gsn-vvc}

    \footnotesize
	\begin{tabular}{p{0.30\textwidth}p{0.22\textwidth}p{0.34\textwidth}c}
		\\
		\hline
		{\raggedright Network\par} & {\raggedright $(d_{in}, d_{lstm})$\par} & {\raggedright Input / layer\par} & $R^2$\\
		\hline\hline
		\multicolumn{4}{l}{\textit{GPNs}}\\
		\hline
		{\raggedright GPN-B-ILSVRC\par} & {\raggedright 0.1, 0.1\par} & {\raggedright glimpse 0, lstm\_\allowbreak out\par} & 0.16\\
		\hline
		{\raggedright GPN-S-ILSVRC\par} & {\raggedright 0.1, 0.1\par} & {\raggedright glimpse 0, lstm\_\allowbreak out\par} & 0.13\\
		\hline
		{\raggedright GPN-R-ILSVRC\par} & {\raggedright 0.25, 0.25\par} & {\raggedright glimpse 4, lstm\_\allowbreak out\par} & 0.18\\
		\hline
		{\raggedright GPN-RS-ILSVRC\par} & {\raggedright 0.1, 0.1\par} & {\raggedright glimpse 4, lstm\_\allowbreak out\par} & 0.16\\
		\hline
		{\raggedright GPN-B-SimCLR\par} & {\raggedright 0.1, 0.1\par} & {\raggedright glimpse 0, lstm\_\allowbreak out\par} & 0.22\\
		\hline
		{\raggedright GPN-S-SimCLR\par} & {\raggedright 0.1, 0.1\par} & {\raggedright glimpse 3, lstm\_\allowbreak out\par} & 0.16\\
		\hline
		{\raggedright GPN-R-SimCLR\par} & {\raggedright 0.25, 0.1\par} & {\raggedright glimpse 5, lstm\_\allowbreak out\par} & 0.23\\
		\hline
		{\raggedright GPN-RS-SimCLR\par} & {\raggedright 0.1, 0.1\par} & {\raggedright glimpse 4, lstm\_\allowbreak out\par} & 0.20\\
		\hline
		{\raggedright GPN-B-Dinov2B\par} & {\raggedright 0.1, 0.1\par} & {\raggedright glimpse 0, lstm\_\allowbreak out\par} & 0.13\\
		\hline
		{\raggedright GPN-S-Dinov2B\par} & {\raggedright 0.1, 0.1\par} & {\raggedright glimpse 3, lstm\_\allowbreak out\par} & 0.10\\
		\hline
		{\raggedright GPN-R-Dinov2B\par} & {\raggedright 0.1, 0.1\par} & {\raggedright glimpse 4, lstm\_\allowbreak out\par} & 0.14\\
		\hline
		{\raggedright GPN-RS-Dinov2B\par} & {\raggedright 0.1, 0.1\par} & {\raggedright glimpse 2, lstm\_\allowbreak out\par} & 0.15\\
		\hline\hline
		\multicolumn{4}{l}{\textit{GSNs}}\\
		\hline
		{\raggedright cGSN-R-SimCLR\par} & {\raggedright 0.5, 0.25\par} & {\raggedright glimpse 0, joint\_\allowbreak proj\par} & 0.14\\
		\hline
		{\raggedright sGSN-R-SimCLR\par} & {\raggedright 0.5, 0.5\par} & {\raggedright glimpse 0, joint\_\allowbreak proj\par} & 0.16\\
		\hline
	\end{tabular}
\end{table}

\begin{table}
	\centering
	\caption{\textbf{DINO-family self-supervised vision models tested for VVC alignment.}
		For each model, the table lists the source, the best-aligned input/layer, and the average VVC alignment. TH refers to PyTorch Hub; HF refers to Hugging Face.}
	\label{tab:vision-ssl-dino-vvc}

    \footnotesize
	\begin{tabular}{p{0.24\textwidth}p{0.30\textwidth}p{0.34\textwidth}c}
		\\
		\hline
		{\raggedright Network\par} & {\raggedright Source\par} & {\raggedright Input / layer\par} & $R^2$\\
		\hline\hline
		{\raggedright dinov2\_\allowbreak vitb14 (DINOv2-B)\par} & {\raggedright TH: facebookresearch/\allowbreak dinov2\par} & {\raggedright central glimpse, blocks.\allowbreak 11.\allowbreak mlp.\allowbreak fc1\par} & 0.15\\
		\hline
		{\raggedright dinov2\_\allowbreak vitl14\par} & {\raggedright TH: facebookresearch/\allowbreak dinov2\par} & {\raggedright central glimpse, blocks.\allowbreak 23.\allowbreak norm2\par} & 0.15\\
		\hline
		{\raggedright dinov2\_\allowbreak vitg14\_\allowbreak reg\par} & {\raggedright TH: facebookresearch/\allowbreak dinov2\par} & {\raggedright central glimpse, blocks.\allowbreak 22.\allowbreak norm2\par} & 0.10\\
		\hline
		{\raggedright dino\_\allowbreak webssl\_\allowbreak 1b\par} & {\raggedright HF: facebook/\allowbreak webssl-dino1b-\allowbreak full2b-\allowbreak 224\par} & {\raggedright central glimpse, encoder.\allowbreak layer.\allowbreak 21.\allowbreak mlp.\allowbreak weights\_\allowbreak in\par} & 0.12\\
		\hline
		{\raggedright dino\_\allowbreak webssl\_\allowbreak 3b\par} & {\raggedright HF: facebook/\allowbreak webssl-dino3b-\allowbreak full2b-\allowbreak 224\par} & {\raggedright central glimpse, encoder.\allowbreak layer.\allowbreak 16.\allowbreak norm2\par} & 0.12\\
		\hline
		{\raggedright dino\_\allowbreak webssl\_\allowbreak 7b\par} & {\raggedright HF: facebook/\allowbreak webssl-dino7b-\allowbreak full8b-\allowbreak 518\par} & {\raggedright central glimpse, encoder.\allowbreak layer.\allowbreak 20.\allowbreak norm2\par} & 0.11\\
		\hline
		{\raggedright dinov3\_\allowbreak vitb16\par} & {\raggedright HF: facebook/\allowbreak dinov3-vitb16-\allowbreak pretrain-\allowbreak lvd1689m\par} & {\raggedright central glimpse, layer.\allowbreak 8.\allowbreak norm2\par} & 0.13\\
		\hline
		{\raggedright dinov3\_\allowbreak convnext\_\allowbreak large\par} & {\raggedright HF: facebook/\allowbreak dinov3-convnext-\allowbreak large-pretrain-\allowbreak lvd1689m\par} & {\raggedright central glimpse, stages.\allowbreak 2.\allowbreak layers.\allowbreak 20.\allowbreak depthwise\_\allowbreak conv\par} & 0.15\\
		\hline
		{\raggedright dinov3\_\allowbreak vit7b16\par} & {\raggedright HF: facebook/\allowbreak dinov3-vit7b16-\allowbreak pretrain-\allowbreak lvd1689m\par} & {\raggedright central glimpse, layer.\allowbreak 21.\allowbreak mlp.\allowbreak act\_\allowbreak fn\par} & 0.12\\
		\hline
	\end{tabular}
\end{table}

\begin{table}
	\centering
	\caption{\textbf{iJEPA and MAE self-supervised vision models tested for VVC alignment.}
		For each model, the table lists the source, the best-aligned input/layer, and the average VVC alignment. HF refers to Hugging Face.}
	\label{tab:vision-ssl-ijepa-mae-vvc}

    \footnotesize
	\begin{tabular}{p{0.24\textwidth}p{0.30\textwidth}p{0.34\textwidth}c}
		\\
		\hline
		{\raggedright Network\par} & {\raggedright Source\par} & {\raggedright Input / layer\par} & $R^2$\\
		\hline\hline
		\multicolumn{4}{l}{\textit{iJEPA}}\\
		\hline
		{\raggedright ijepa\_\allowbreak vith14\_\allowbreak 1k\par} & {\raggedright HF: jmtzt/\allowbreak ijepa\_\allowbreak vith14\_\allowbreak 1k\par} & {\raggedright full scene, encoder.\allowbreak layer.\allowbreak 24.\allowbreak intermediate.\allowbreak dense\par} & 0.10\\
		\hline
		{\raggedright ijepa\_\allowbreak vitg16\_\allowbreak 22k\par} & {\raggedright HF: jmtzt/\allowbreak ijepa\_\allowbreak vitg16\_\allowbreak 22k\par} & {\raggedright central glimpse, encoder.\allowbreak layer.\allowbreak 31.\allowbreak intermediate.\allowbreak dense\par} & 0.11\\
		\hline\hline
		\multicolumn{4}{l}{\textit{MAE}}\\
		\hline
		{\raggedright mae\_\allowbreak vitb\par} & {\raggedright HF: facebook/\allowbreak vit-mae-\allowbreak base\par} & {\raggedright central glimpse, encoder.\allowbreak layer.\allowbreak 8.\allowbreak intermediate.\allowbreak dense\par} & 0.08\\
		\hline
		{\raggedright mae\_\allowbreak vitl\par} & {\raggedright HF: facebook/\allowbreak vit-mae-\allowbreak large\par} & {\raggedright central glimpse, encoder.\allowbreak layer.\allowbreak 14.\allowbreak intermediate.\allowbreak dense\par} & 0.09\\
		\hline
		{\raggedright mae\_\allowbreak vith\par} & {\raggedright HF: facebook/\allowbreak vit-mae-\allowbreak huge\par} & {\raggedright central glimpse, encoder.\allowbreak layer.\allowbreak 15.\allowbreak intermediate.\allowbreak dense\par} & 0.11\\
		\hline
		{\raggedright mae\_\allowbreak webssl\_\allowbreak 300m\par} & {\raggedright HF: facebook/\allowbreak webssl-mae300m-\allowbreak full2b-\allowbreak 224\par} & {\raggedright full scene, encoder.\allowbreak layer.\allowbreak 14.\allowbreak attention.\allowbreak attention\par} & 0.07\\
		\hline
		{\raggedright mae\_\allowbreak webssl\_\allowbreak 1b\par} & {\raggedright HF: facebook/\allowbreak webssl-mae1b-\allowbreak full2b-\allowbreak 224\par} & {\raggedright central glimpse, encoder.\allowbreak layer.\allowbreak 21.\allowbreak intermediate.\allowbreak dense\par} & 0.10\\
		\hline
		{\raggedright mae\_\allowbreak webssl\_\allowbreak 3b\par} & {\raggedright HF: facebook/\allowbreak webssl-mae3b-\allowbreak full2b-\allowbreak 224\par} & {\raggedright central glimpse, encoder.\allowbreak layer.\allowbreak 14.\allowbreak intermediate.\allowbreak dense\par} & 0.08\\
		\hline
	\end{tabular}
\end{table}

\begin{table}
	\centering
	\caption{\textbf{Vision-only supervised and architecture-optimized models tested for VVC alignment.}
		For each model, the table lists the source, the best-aligned input/layer, and the average VVC alignment.}
	\label{tab:vision-sota-vvc}

    \footnotesize
	\begin{tabular}{p{0.24\textwidth}p{0.30\textwidth}p{0.34\textwidth}c}
		\\
		\hline
		{\raggedright Network\par} & {\raggedright Source\par} & {\raggedright Input / layer\par} & $R^2$\\
		\hline\hline
		\multicolumn{4}{l}{\textit{EfficientNet}}\\
		\hline
		{\raggedright efficientnet\_\allowbreak b1\par} & {\raggedright torchvision models\par} & {\raggedright central glimpse, features.\allowbreak 6.\allowbreak 3.\allowbreak block.\allowbreak 2.\allowbreak fc1\par} & 0.18\\
		\hline
		{\raggedright efficientnet\_\allowbreak b3\par} & {\raggedright torchvision models\par} & {\raggedright central glimpse, features.\allowbreak 6.\allowbreak 4.\allowbreak block.\allowbreak 2.\allowbreak fc1\par} & 0.19\\
		\hline
		{\raggedright efficientnet\_\allowbreak b7\par} & {\raggedright torchvision models\par} & {\raggedright central glimpse, features.\allowbreak 6.\allowbreak 6.\allowbreak block.\allowbreak 2.\allowbreak fc1\par} & 0.17\\
		\hline\hline
		\multicolumn{4}{l}{\textit{LeViT}}\\
		\hline
		{\raggedright levit\_\allowbreak 128\par} & {\raggedright timm: levit\_\allowbreak 128.\allowbreak fb\_\allowbreak dist\_\allowbreak in1k\par} & {\raggedright central glimpse, stages.\allowbreak 2.\allowbreak blocks.\allowbreak 0.\allowbreak attn.\allowbreak qkv.\allowbreak linear\par} & 0.18\\
		\hline
		{\raggedright levit\_\allowbreak 256\par} & {\raggedright timm: levit\_\allowbreak 256.\allowbreak fb\_\allowbreak dist\_\allowbreak in1k\par} & {\raggedright full scene, stages.\allowbreak 1.\allowbreak blocks.\allowbreak 0.\allowbreak attn\par} & 0.13\\
		\hline
		{\raggedright levit\_\allowbreak 384\par} & {\raggedright timm: levit\_\allowbreak 384.\allowbreak fb\_\allowbreak dist\_\allowbreak in1k\par} & {\raggedright central glimpse, stages.\allowbreak 1.\allowbreak blocks.\allowbreak 0.\allowbreak attn\par} & 0.13\\
		\hline\hline
		\multicolumn{4}{l}{\textit{HardCoreNAS}}\\
		\hline
		{\raggedright hardcorenas\_\allowbreak a\par} & {\raggedright timm: hardcorenas\_\allowbreak a.\allowbreak miil\_\allowbreak green\_\allowbreak in1k\par} & {\raggedright full scene, blocks.\allowbreak 5.\allowbreak 1.\allowbreak se.\allowbreak conv\_\allowbreak reduce\par} & 0.17\\
		\hline
		{\raggedright hardcorenas\_\allowbreak d\par} & {\raggedright timm: hardcorenas\_\allowbreak d.\allowbreak miil\_\allowbreak green\_\allowbreak in1k\par} & {\raggedright central glimpse, blocks.\allowbreak 5.\allowbreak 1.\allowbreak se.\allowbreak conv\_\allowbreak reduce\par} & 0.18\\
		\hline
		{\raggedright hardcorenas\_\allowbreak f\par} & {\raggedright timm: hardcorenas\_\allowbreak f.\allowbreak miil\_\allowbreak green\_\allowbreak in1k\par} & {\raggedright full scene, blocks.\allowbreak 4.\allowbreak 3.\allowbreak se.\allowbreak act1\par} & 0.17\\
		\hline
	\end{tabular}
\end{table}

\begin{table}
	\centering
	\caption{\textbf{Visual baseline models tested for VVC alignment.}
		For each model, the table lists the source, the best-aligned input/layer, and the average VVC alignment. DG3 denotes DeepGazeIII.}
	\label{tab:visual-baselines-vvc}

    \footnotesize
	\begin{tabular}{p{0.24\textwidth}p{0.30\textwidth}p{0.34\textwidth}c}
		\\
		\hline
		{\raggedright Network\par} & {\raggedright Source\par} & {\raggedright Input / layer\par} & $R^2$\\
		\hline\hline
		{\raggedright RN50-ILSVRC\par} & {\raggedright torchvision models: v1\par} & {\raggedright full scene, layer4.\allowbreak 1.\allowbreak bn2\par} & 0.13\\
		\hline
		{\raggedright RN50-SimCLR\par} & {\raggedright lightly-ai: simclrv1-\allowbreak imagenet1k-\allowbreak resnet50-\allowbreak 1x\par} & {\raggedright full scene, avgpool\par} & 0.12\\
		\hline
		{\raggedright DG3\par} & {\raggedright github: matthias-k/\allowbreak DeepGaze\par} & {\raggedright central glimpse, features.\allowbreak features.\allowbreak 1.\allowbreak features.\allowbreak denseblock3.\allowbreak denselayer28.\allowbreak norm2\par} & 0.14\\
		\hline
	\end{tabular}
\end{table}

\begin{table}
	\centering
	\caption{\textbf{Vision-language and semantic baseline models tested for VVC alignment.}
		For each model, the table lists the source, the best-aligned input/layer, and the average VVC alignment. HF refers to Hugging Face.}
	\label{tab:vision-language-vvc}

    \footnotesize
	\begin{tabular}{p{0.24\textwidth}p{0.30\textwidth}p{0.34\textwidth}c}
		\\
		\hline
		{\raggedright Network\par} & {\raggedright Source\par} & {\raggedright Input / layer\par} & $R^2$\\
		\hline\hline
		\multicolumn{4}{l}{\textit{CLIP}}\\
		\hline
		{\raggedright clip\_\allowbreak RN50\par} & {\raggedright github: openai/CLIP: RN50\par} & {\raggedright full scene, visual.\allowbreak layer4.\allowbreak 1.\allowbreak bn1\par} & 0.11\\
		\hline
		{\raggedright clip\_\allowbreak vitb\par} & {\raggedright github: openai/CLIP: ViT-B/16\par} & {\raggedright central glimpse, visual.\allowbreak transformer.\allowbreak resblocks.\allowbreak 8.\allowbreak ln\_\allowbreak 2\par} & 0.13\\
		\hline
		{\raggedright clip\_\allowbreak vitl\par} & {\raggedright github: openai/CLIP: ViT-L/14@336px\par} & {\raggedright central glimpse, visual.\allowbreak transformer.\allowbreak resblocks.\allowbreak 11.\allowbreak attn\par} & 0.09\\
		\hline\hline
		\multicolumn{4}{l}{\textit{SigLIP}}\\
		\hline
		{\raggedright siglip2\_\allowbreak b\par} & {\raggedright HF: google/\allowbreak siglip2-base-\allowbreak patch16-\allowbreak 224\par} & {\raggedright central glimpse, vision\_\allowbreak model.\allowbreak encoder.\allowbreak layers.\allowbreak 7.\allowbreak layer\_\allowbreak norm2\par} & 0.17\\
		\hline
		{\raggedright siglip2\_\allowbreak l\par} & {\raggedright HF: google/\allowbreak siglip2-large-\allowbreak patch16-\allowbreak 256\par} & {\raggedright central glimpse, vision\_\allowbreak model.\allowbreak encoder.\allowbreak layers.\allowbreak 12.\allowbreak mlp.\allowbreak fc1\par} & 0.14\\
		\hline
		{\raggedright siglip2\_\allowbreak g\par} & {\raggedright HF: google/\allowbreak siglip2-giant-\allowbreak opt-patch16-\allowbreak 384\par} & {\raggedright central glimpse, vision\_\allowbreak model.\allowbreak encoder.\allowbreak layers.\allowbreak 15.\allowbreak mlp.\allowbreak fc1\par} & 0.13\\
		\hline\hline
		\multicolumn{4}{l}{\textit{Semantic baseline}}\\
		\hline
		{\raggedright BLT\_\allowbreak MPNet\par} & {\raggedright Doerig et al. 2025 \par} & {\raggedright full scene, layernorm\_\allowbreak layer\_\allowbreak 9\_\allowbreak time\_\allowbreak 0\par} & 0.13\\
		\hline
	\end{tabular}
\end{table}

\end{document}